# A Real-Time Approach to Autonomous CAN Bus Reverse Engineering

Kevin Setterstrom
Department of Computer Science
North Dakota State University
1320 Albrecht Blvd., Room 258
Fargo, ND 58108
P: +1 (701) 231-8562
F: +1 (701) 231-8255
E: kevin.setterstrom@ndsu.edu

Jeremy Straub[1]
Center for Cybersecurity and AI
University of West Florida
220 W. Garden St., Suite 250
Pensacola, FL 32502
P: +1 (850) 474-2999
E: jstraub@uwf.edu

**Abstract**

This paper introduces a real-time method for reverse engineering a vehicle's CAN bus without prior knowledge of the vehicle or its CAN system. By comparing inertial measurement and CAN data during significant vehicle events, the method accurately identified the CAN channels associated with the accelerator pedal, brake pedal, and steering wheel. Utilizing an IMU, CAN module, and event-driven software architecture, the system was validated using prerecorded serialized data from previous studies. This data, collected during multiple vehicle drives, included synchronized IMU and CAN recordings. By using these consistent datasets, the improvements made in this work were tested and validated under the same conditions as in the previous studies, enabling direct comparison to earlier results. Faster processing times were produced and less computational power was needed, as compared to the earlier methods. This work could have potential application to making aftermarket autonomous vehicle kits and for cybersecurity applications. It is a scalable and adaptable solution for autonomous CAN reverse engineering in near real-time.

## 1. Introduction

Self-driving cars are at the forefront of today's technological advancements. Thanks to the controller area network (CAN), vehicles now come equipped with features such as autonomous emergency braking (AEB), lane keep assist (LKAS), and adaptive cruise control (ACC), which enhance both performance and safety [1], [2]. While these capabilities improve safety and performance, they also introduce significant complexity in to vehicle communication architectures. This complexity presents a major challenge for the development of universal aftermarket autonomy and diagnostics solutions, as CAN message definitions and signal layouts are highly vehicle – and OEM – specific.

---
[1] This work was partially completed while J. Straub was at North Dakota State University.

In response to this challenge, several companies [3], [4] have developed aftermarket self-driving platforms that interface directly with vehicle CAN buses. However, widespread adoption remains limited because each supported vehicle requires extensive manual reverse engineering of proprietary CAN messages. As a result, aftermarket solution scalability across vehicle models and manufacturers remains a barrier.

Recent research has demonstrated that it is possible to reverse engineer the CAN messages associated with vehicle control inputs without prior knowledge of a vehicle's internal CAN architecture [5], [6]. These approaches rely on correlating observed vehicle behavior with CAN traffic to infer control-related signals. While effective, the existing methods have largely focused on offline analysis and have not been designed to operate under real-time constraints, limiting their applicability to live vehicle systems.

This paper addresses this limitation by introducing a real-time method for reverse engineering the automotive CAN channels associated with key vehicle controls, specifically the accelerator pedal input, the brake pedal input, and steering wheel motion, without requiring any prior knowledge of the vehicle or its CAN message definitions. The proposed approach simultaneously captures inertial measurement data and CAN traffic during discrete vehicle events – such as acceleration, braking, and steering – enabling semantic inference of control-related CAN channels during live operation.

Building upon prior work [5], [6], the system uses the Robot Operating System (ROS) and its native rosbag functionality [7] to enable direct comparison between the offline and real-time reverse engineering approaches. By replaying the same datasets used in previous studies, this work provides a controlled evaluation of real-time performance while ensuring consistency with the established results. This comparison highlights the feasibility of transitioning CAN reverse engineering from an offline process to a real-time capability.

The remainder of this paper is organized as follows. Section 2 reviews relevant background and prior research. Section 3 presents the system architecture and methodology, including the hardware, software, and real-time enhancements. Section 4 reports the experimental results obtained across multiple vehicle platforms. Section 5 discusses these findings in the context of prior work, system robustness, and external validation. Finally, Section 6 concludes the paper and outlines directions for future research.

## 2. Background

This section reviews foundational concepts and prior research relevant to real-time CAN bus reverse engineering. It begins by outlining the general phases involved in CAN reverse engineering. This is followed by a summary of prior studies that directly inform the methodology used in this work. The section concludes with an overview of existing CAN channel translation approaches in the literature, providing context for the design choices made in this paper.

## 2.1. Introduction to Controller Area Network Reverse Engineering

CAN reverse engineering is commonly described as consisting of two complementary phases: tokenization and translation [8]. Tokenization refers to the identification of signal boundaries within CAN frames, including bit-level start positions, lengths, and encoding formats. Translation, in contrast, involves assigning semantic meaning to candidate signals by associating them with vehicle states or control actions, such as accelerator input, braking, or steering.

While precise tokenization is required for full OEM-level understanding of CAN message structure, this study does not attempt to recover true signal boundaries. Instead, it uses fixed-width channel hypotheses that serve as candidate representations for semantic inference. This allows the system to focus on translation rather than structural recovery, enabling real-time operation without prior vehicle knowledge.

Accordingly, the contribution of this work lies in the translation phase, where candidate channels are evaluated and ranked based on their semantic alignment with observed vehicle behavior. Tokenization is treated as an orthogonal problem that can be integrated into future systems to achieve full end-to-end CAN reverse engineering.

## 2.2. Summary of Previous Work

This research builds directly on two prior studies that explored autonomous CAN bus reverse engineering without prior vehicle knowledge.

The first study [5] introduced a system that was capable of identifying the CAN channels associated with accelerator and brake pedal inputs across multiple vehicles. The system combined inertial measurement data with CAN data collected from the vehicle's high-speed CAN bus via the OBD-II interface. Correlation analysis between the physical pedal inputs and the CAN signals enabled accurate channel identification through offline post-processing.

While effective, this study revealed a limitation in correlating brake pedal input during periods when the vehicle was stationary, as inertial data alone was insufficient to disambiguate braking events under those conditions. This limitation motivated the incorporation of additional data sources.

The second study [6] extended this approach by integrating Global Positioning System (GPS) data to provide direct measurements of vehicle velocity. By identifying stationary intervals using GPS, the system was able to exclude misleading deceleration data and significantly improve brake-related signal inference. This enhancement also enabled the successful reverse engineering of steering wheel position signals, expanding the scope of identifiable vehicle control signals.

Together, these studies demonstrated that CAN channel translation can be achieved without prior vehicle knowledge through behavioral correlation. However, both approaches used offline analysis and were not designed to operate under real-time constraints, limiting their applicability to live vehicle systems.

While this prior work established the feasibility of autonomous CAN channel translation, neither study demonstrated the ability to perform this process in real time. Offline processing requires full drive recordings, extensive post-processing, and nontrivial computational resources, which restricts deployment in lightweight or embedded systems.

A real-time CAN translation capability enables a fundamentally different class of applications. By producing actionable results during vehicle operation, such a system can support modular aftermarket autonomy, adaptive diagnostics, and live cybersecurity monitoring without requiring extensive vehicle-specific configuration. These requirements motivate the development of a real-time translation framework capable of operating autonomously and efficiently under live conditions.

## 2.3. Expansion on Work in the Field

Existing CAN channel translation techniques can be broadly grouped into four categories: parameter identifier-based methods, machine learning approaches, frame matching techniques, and semantic taxonomy-based methods [8].

Parameter identifier-based approaches use OBD-II diagnostic queries to obtain standardized vehicle measurements and correlate them with raw CAN traffic. Kang, et al. [9] introduced an automated framework that repeatedly queries OBD-II parameter identifiers and searches for matching byte patterns in CAN frames. It refines candidates across multiple interrogations and validates results through message injection.

Blaauwendraad and Kieberl [10] extended this idea by using Pearson correlation coefficients rather than direct byte matching. This enabled the detection of CAN signals, even when manufacturers apply linear scaling or translation to the stored values.

Similarly, Huybrechts, et al. [11] combined OBD-II–derived reference signals with statistical comparison techniques, such as RMS deviation, and machine learning classifiers to automate signal identification. While parameter identifiers are useful for extracting standardized vehicle data, they limit systems to using the available parameter identifiers and may not identify all control-related CAN signals. This reduces their applicability for reverse engineering [9].

Machine learning has been used for CAN reverse engineering, particularly for identifying signals and understanding vehicle behavior from raw traffic. Jaynes, et al. [12] approached it as a supervised classification task. They trained models – such as k-nearest neighbor, multilayer perceptron, and random forest – directly with byte-level payload data to associate messages with ECU functions.

Buscemi, et al. [13] built on this idea by capturing how signals behave over time, such as whether they are static, dynamic, or counter-like. They then applied logical constraints to ensure that the final predictions were consistent with known vehicle structures.

Moore, et al. [14] focused on the vehicle state level, rather than individual signal decoding. They identified speed-related signals, through statistical correlation, and modeled driver states using hidden Markov models and convolutional neural networks that were trained on time-window representations of CAN activity. In contrast, Ezeobi, et al. [15] used unsupervised clustering methods to group CAN messages based on temporal patterns. This enabled functional grouping, without relying on databases or manual labeling.

Frame matching approaches leverage previously decoded vehicles as reference datasets to accelerate reverse engineering on new platforms. CANMatch [16] exploits the reuse of CAN frame identifiers across vehicle models by using a database of frame IDs to identify frame IDs in a new vehicle. When a match is identified, previously extracted signal boundaries and semantics are transferred directly, reducing the manual decoding effort required. While effective when identifiers are reused, this approach depends heavily on access to reference datasets. This limits its applicability in some environments.

Semantic taxonomy-based reverse engineering approaches infer CAN message meaning by correlating observable vehicle behavior with underlying bus activity, rather than relying on protocol structure alone. Signals are interpreted through externally measurable actions, such as vehicle motion, driver interaction, or environmental context.

Frassinelli, et al. [17] demonstrated that physical vehicle trajectories, reconstructed from GPS and motion data, could be matched with CAN traffic patterns to infer sensitive information about vehicle location and behavior. This linked bus-level signals to real-world semantics. Similarly, LibreCAN [18] automated large-scale message translation by extracting candidate signals from raw CAN traffic and correlating them with openly available OBD-II parameters and smartphone inertial measurements to identify kinematic functions. In both cases, semantic meaning is derived through behavioral alignment rather than through formal protocol decoding. These approaches map observable system dynamics to CAN traffic patterns, grouping signals by their functional contribution to vehicle behavior.

The methodology presented in this paper builds upon the semantic taxonomy paradigm, extending it to operate under real-time constraints. While tokenization is not addressed in this work, the proposed translation framework is compatible with systems that combine structural and semantic inference to achieve complete CAN reverse engineering.

## 3. Methodology

This section describes the methodology that was used to achieve real-time reverse engineering of vehicle CAN messages during vehicle operation. Building on prior work [5], [6], the proposed approach introduces architectural and execution-level enhancements that improve processing efficiency, reduce hardware requirements, and support real-time scalability. Previously recorded rosbag datasets from [5], [6] were used to validate the system under controlled and repeatable conditions.

### 3.1. System Overview

The system used for this research retains the core hardware components used in the prior studies [5], [6], ensuring the direct comparability of results. The platform consists of a laptop, a CAN interface connected to the vehicle's OBD-II port, and a fixed-position inertial measurement unit (IMU), as shown in Figure 1. Together, these components form the foundation for real-time CAN message reverse engineering.

A high-precision IMU [19] equipped with a 3-axis gyroscope and 3-axis accelerometer provides measurements of linear acceleration and angular velocity. These measurements are used to detect vehicle motion events associated with acceleration, braking, and steering.

CAN traffic is acquired using a CAN-USB analyzer [20] connected to the vehicle's high-speed CAN bus via the OBD-II interface. This module captures raw CAN frames during vehicle operations and transmits them to the processing system.

A Dell Precision 5530 laptop running Ubuntu 20.04 Linux was used as the central processing and operator interface. The system hosts the software components responsible for data acquisition, real-time inference, and visualization of reverse-engineered CAN channels.

The Robot Operating System (ROS) was selected as the middleware framework, due to its support for distributed processing, time-synchronized data exchange, and modular system design [21]. ROS also provides the rosbag utility [7], which enables recording and replay of synchronized sensor and CAN data. Previously recorded rosbags from [5], [6] were reused to validate the real-time system under identical conditions, eliminating variability associated with live testing while preserving temporal fidelity.

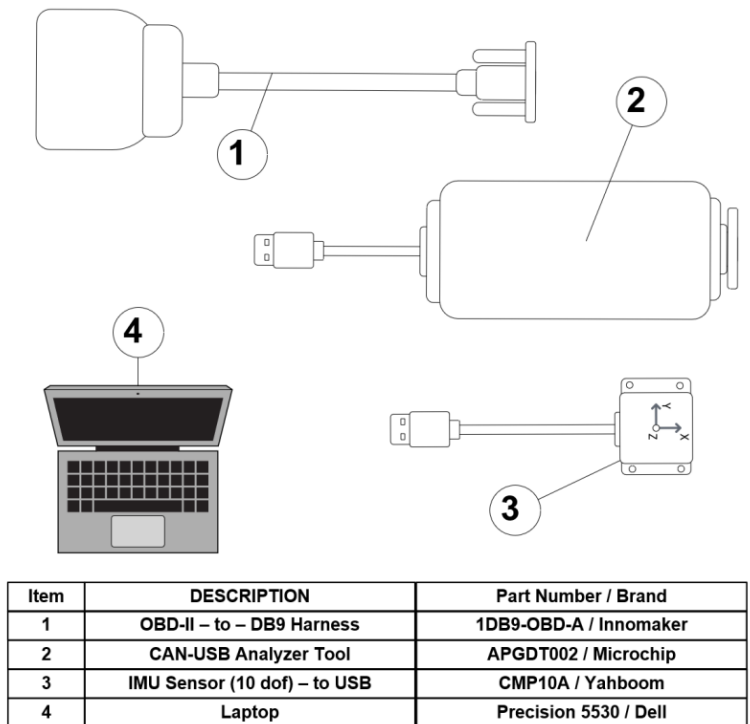

| Item | DESCRIPTION | Part Number / Brand |
|---|---|---|
| 1 | OBD-II – to – DB9 Harness | 1DB9-OBD-A / Innomaker |
| 2 | CAN-USB Analyzer Tool | APGDT002 / Microchip |
| 3 | IMU Sensor (10 dof) – to USB | CMP10A / Yahboom |
| 4 | Laptop | Precision 5530 / Dell |

**Figure 1.** Real-time automotive CAN reverse engineering equipment.

*3.2. Key Methodological Enhancements*

This section introduces several key enhancements to the methodology, addressing challenges identified in earlier studies. The enhancements focus on improving efficiency, accuracy, and scalability through the adoption of an event-based architecture, hardware simplification, multithreading, CAN channel masking, and changes to the tokenization process. Each is now discussed.

*3.2.1. Event-Based Architecture*

A central enhancement introduced in this work is the adoption of an event-based architecture for real-time data collection and analysis. Rather than processing complete CAN recordings, the system focuses on temporally localized vehicle events such as acceleration, deceleration, and steering. By constraining analysis to these events, the system reduces computational load while improving the signal-to-noise ratio for semantic inference.

The architecture is inherently event-driven: detected vehicle actions initiate processing across successive layers of the system. Each layer refines and contextualizes event-related data before passing it downstream, enabling real-time operation while preserving semantic relevance.

The use of rosbag replay allows these event-driven mechanisms to be evaluated against identical datasets used in prior work [5], [6], ensuring that improvements can be attributed to methodological changes rather than data variability.

As illustrated in Figure 2, the event-based software architecture consists of five layers. Each layer is responsible for a specific stage of data acquisition, interpretation, and validation. The layers are now discussed.

**Layer 1: Measurement Layer** – The architecture begins is the measurement layer**,** which comprised of the IMU node and the CAN node. These nodes are responsible for real-time data acquisition. The IMU captures linear acceleration and angular velocity, while the CAN node interfaces with the vehicle's CAN bus to collect raw CAN frames. These measurements provide the inputs required for detecting and characterizing vehicle motion events.

**Layer 2: Event Detection Layer** – The event detection layer processes the real-time data streams from the measurement layer to identify discrete vehicle actions, including acceleration, deceleration, and steering. This layer consists of dedicated event detection nodes for each action type which monitor sensor data and identify event-specific time windows.

When an event is detected, the corresponding node captures data during the event and within a bounded temporal window before and after the event. This windowed capture ensures that the system preserves contextual information associated with event onset and completion. The window duration is intentionally constrained to maintain system responsiveness and limit the inclusion of unrelated data, supporting efficient real-time processing.

**Layer 3: Correlation Layer** – The correlation layer establishes relationships between detected vehicle events and candidate CAN channels. Separate correlation nodes analyze acceleration, deceleration, and steering events to identify CAN channels whose temporal behavior most closely aligns with the corresponding vehicle motion.

This layer employs the rate-of-change correlation algorithm introduced in [5] and [6]. It has been adapted for real-time execution. By operating on event-localized data, the correlation process is both more efficient and more robust to background CAN traffic, enabling reliable identification of semantically relevant channels.

**Layer 4: Control Discovery Layer** – Using the ranked correlations produced in the previous layer, the control discovery layer evaluates candidate channels to determine which are most likely associated with specific vehicle control inputs. Dedicated discovery nodes for the accelerator pedal, brake pedal, and steering wheel apply the vehicle controls discovery algorithm from [5] and [6].

This layer refines correlation results into actionable control hypotheses, assuming that the vehicle under test exposes CAN messages corresponding to controls' position or state. The output of this layer consists of a small set of high-confidence control-related channels suitable for real-time validation.

**Layer 5: Verification Layer** – The final stage, the verification layer, performs real-time validation of the discovered control channels and provides a graphical user interface (GUI). The GUI enables the operator to interact with the system and visually confirm that inferred CAN channels respond appropriately to physical control inputs.

As the operator actuates vehicle controls, the GUI provides immediate feedback, allowing discovered channels to be verified against observable vehicle behavior. This layer closes the loop between inference and validation, ensuring that semantic interpretations remain consistent with real-world operation.

This layered architecture, which is shown in Figure 2, reduces the volume of data processed while using event context to improve the accuracy and efficiency of CAN message correlation and control discovery.

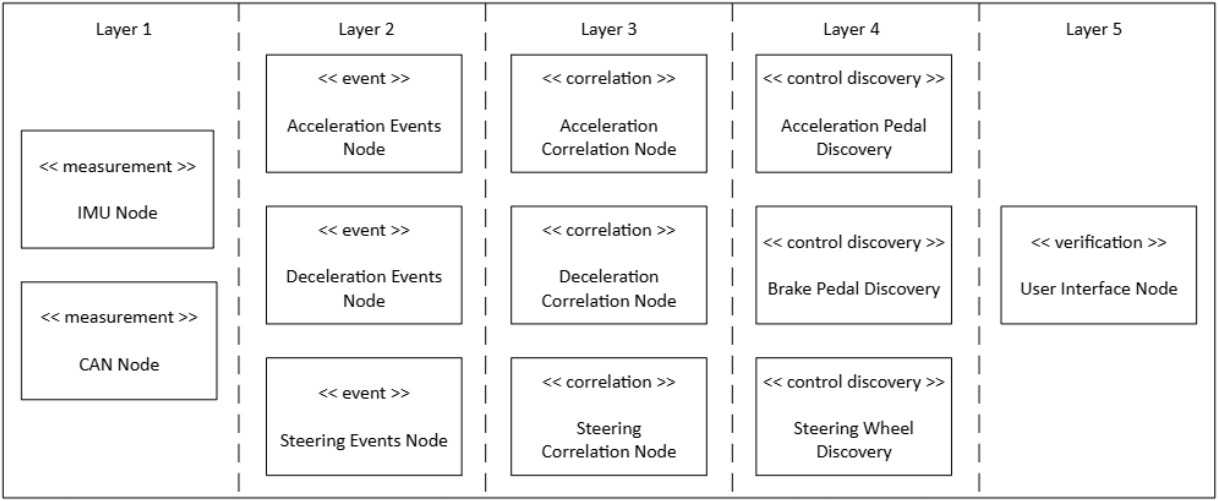

**Figure 2.** Event-based software architecture.

### 3.2.2. Hardware Simplification

Another methodological change introduced in this study is the removal of the GPS sensor, which was used in [6] to identify vehicle standstill conditions and filter out irrelevant deceleration data. By adopting an event-driven recording and analysis strategy, the proposed system evaluates only those time intervals during which meaningful changes in vehicle motion occur. As a result, detection of vehicle rest states is no longer required.

This eliminates the need for a GPS sensor, reducing hardware complexity without sacrificing inference accuracy. Rather than relying on continuous velocity estimation, the system focuses on correlating CAN messages directly with temporally localized motion events, such as acceleration, braking, and steering transitions.

Although background CAN traffic, outside of these events, may still contain useful contextual information, constraining analysis to event-relevant windows improves computational efficiency and strengthens the semantic alignment between the observed vehicle behavior and the candidate CAN channels. This simplification allows the system to have a lightweight and deployable real-time architecture.

### 3.2.3. Multithreading for Improved Processing

To further enhance processing efficiency and scalability, the reverse engineering pipeline is fully integrated within the ROS environment, with each functional component implemented as an independent ROS node executing in its own thread. This multithreaded design enables concurrent processing of measurement acquisition, event detection, correlation analysis, control discovery, and visualization.

ROS employs a publish/subscribe (pub/sub) communication model [21]. This facilitates modularity and scalability by decoupling data producers from consumers. This architecture allows new processing nodes to be added with minimal impact on existing components, supporting extensibility as system complexity increases.

By distributing computation across multiple threads and leveraging ROS's data-centric middleware, the system achieves reduced end-to-end latency and improved responsiveness during real-time operation. This capability is essential for real-time performance as additional event types, control hypotheses, or validation mechanisms are incorporated.

Figure 3 illustrates the ROS-based data-centric middleware architecture. It shows how measurement, event detection, correlation, discovery, and verification nodes communicate through a shared messaging framework.

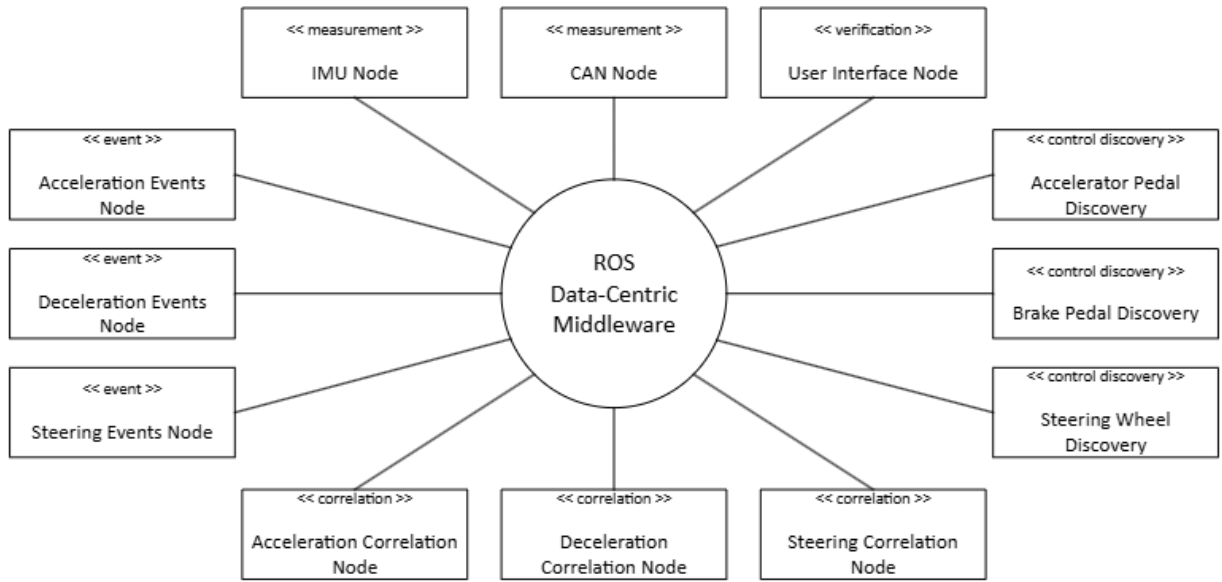

**Figure 3.** ROS data-centric middleware.

### *3.3. CAN Channel Masking and Tokenization Assumption*

To enable real-time operation without prior vehicle knowledge, the proposed system does not attempt to infer true CAN signal boundaries during runtime. Instead, CAN payloads are segmented into uniform, fixed-width channels that serve as candidate signal hypotheses for semantic evaluation.

Payloads are segmented into 16-bit channels. This segmentation choice does not require that underlying vehicle signals are natively 16 bits in length, nor does it represent OEM-level tokenization. Rather, it provides a consistent and computationally tractable abstraction that supports real-time semantic inference across heterogeneous vehicle platforms.

The objective of this segmentation strategy is semantic correlation rather than structural recovery. Even when fixed-width channel boundaries do not align with true signal definitions, semantically meaningful actuator and state correlations can still emerge through time-aligned analysis with inertial and vehicle motion data. This allows the system to identify control-related behavior at the message level without requiring precise knowledge of signal boundaries.

Channel masking is applied in conjunction with fixed-width segmentation to constrain the search space to candidate channels most likely associated with driver control inputs. This masking strategy reduces computational complexity and improves inference robustness while preserving generalizability across vehicles with differing CAN formats.

The selection of a 16-bit channel width was informed by prior work [5], [6] where this granularity proved sufficient to recover semantically meaningful control channels for multiple vehicles. Importantly, the effectiveness of the proposed method does not depend on perfect alignment between the assumed channels and true OEM signals. Instead, the results demonstrate that semantic inference remains robust even under approximate segmentation, reinforcing the generalizability of the approach.

While fixed-width segmentation enables real-time translation in the absence of tokenization, it is not intended as a complete reverse engineering solution. The ultimate goal is to integrate the proposed semantic inference framework with a dedicated CAN tokenization stage that can recover true signal boundaries. Existing tokenization approaches such as READ [22], ACTT [23], LibreCAN [18], CAN-D [24], and CANMatch [16] could be incorporated to achieve full structural and semantic recovery.

By explicitly separating semantic inference from tokenization, this work demonstrates that meaningful CAN channel translation can be achieved in real time without prior vehicle knowledge. This separation enables modular system design and provides a clear path towards future end-to-end CAN reverse engineering that combines real-time inference with precise signal recovery.

### 3.4. Vehicle Control Calibration

The calibration phase provides a controlled reference for validating CAN channels inferred during real-time reverse engineering. Rather than defining true CAN signal boundaries or performing OEM-level decoding, calibration is used to anchor candidate channels to known vehicle control actions, ensuring that subsequent inference focuses on relevant CAN traffic.

Calibration is conducted in a controlled setting to minimize interference from unrelated CAN messages, such as those associated with engine management or background vehicle functions. The procedure mirrors the calibration methods used in prior studies [5], [6]. A key enhancement in this work is the integration of a graphical user interface (GUI). The GUI standardizes the calibration process, facilitates management of multiple vehicle configurations, and streamlines the transition between calibration and real-time inference.

Figure 4 shows the GUI vehicle selection interface, which allows the operator to select and manage vehicle configurations. Figure 5 depicts the calibration selection screen, where existing calibration data may be reused or a new calibration initiated. Figure 6 shows the GUI during a brake pedal calibration, demonstrating the step-by-step guidance provided to the operator.

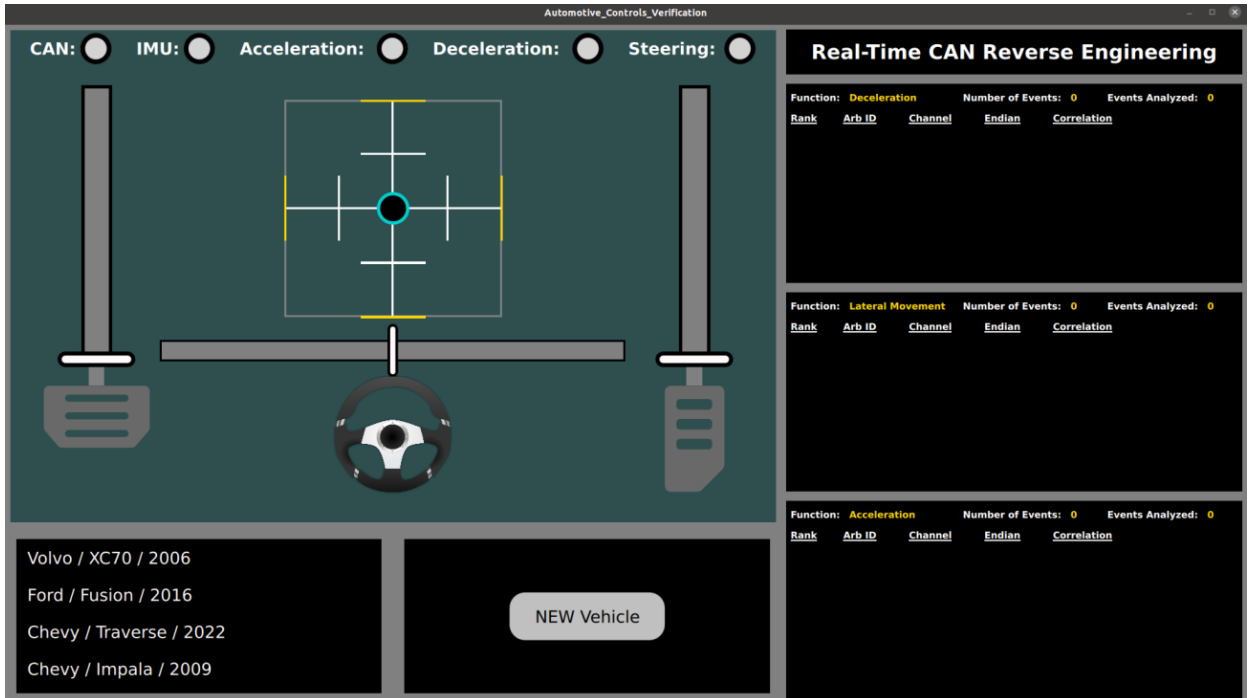

**Figure 4.** GUI display for vehicle selection.

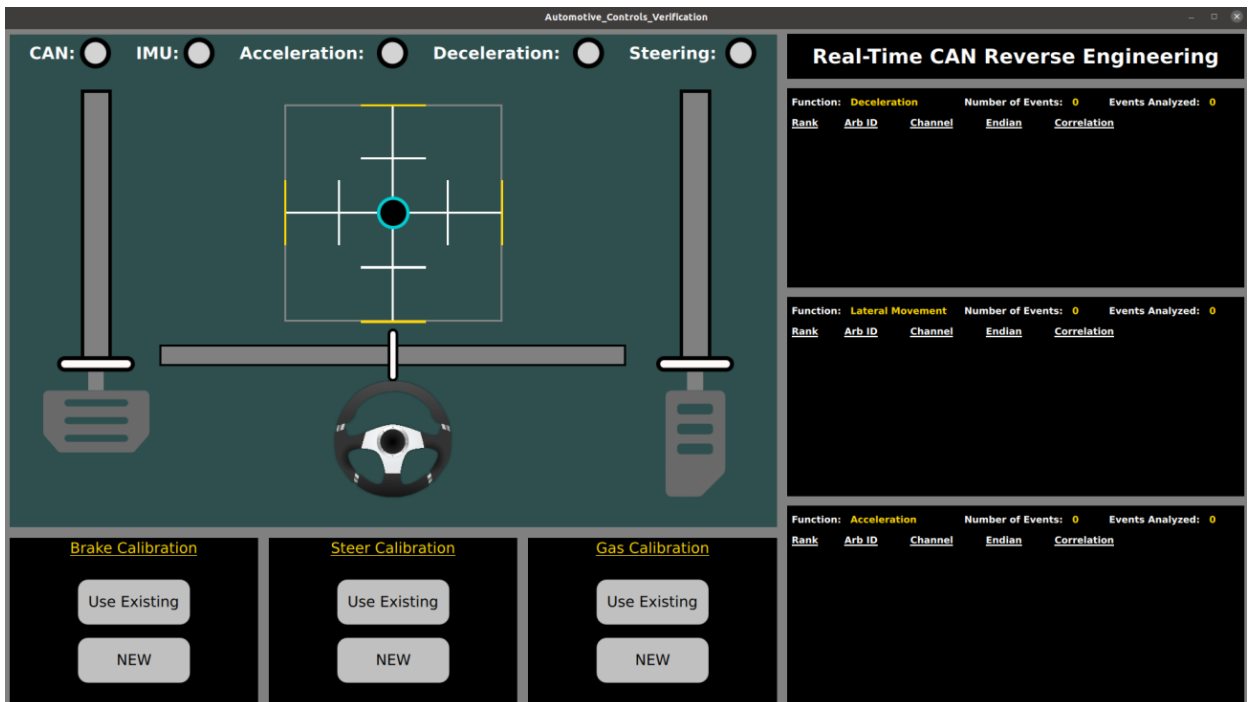

**Figure 5.** GUI display for calibration selection.

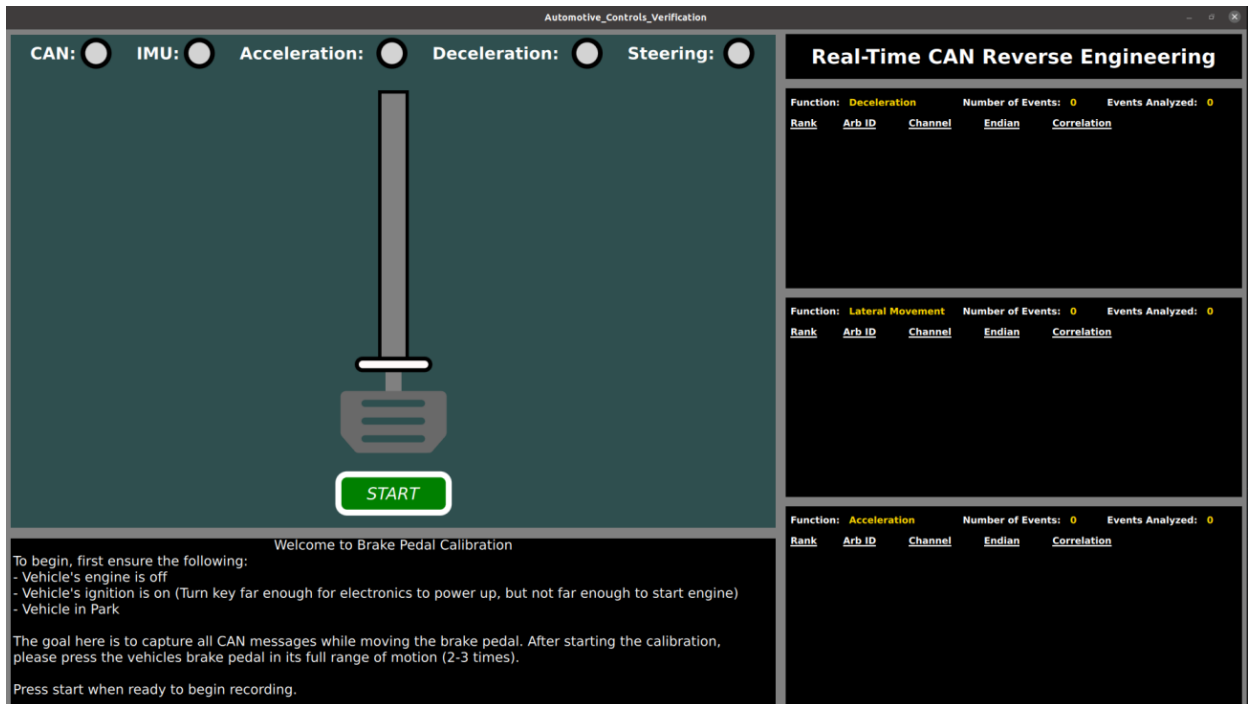

**Figure 6.** GUI display for brake pedal calibration procedure.

The calibration process involves four steps. Each is now discussed.

First, stationary vehicle setup is performed. For the accelerator pedal and brake pedal calibration, the vehicle is placed in park with the ignition on, but the engine is not running. This setup is chosen to eliminate CAN traffic associated with engine operation, ensuring that the signals captured are purely related to the control inputs. For the brake pedal specifically, although the engine is off, pressure built up in the brake lines can affect how far the pedal can be pushed after the first application.

For the steering wheel calibration, the vehicle's engine is kept running to engage the power steering, as it would be extremely difficult to turn the steering wheel without it. This allows for the accurate capture of the CAN signals related to steering input.

Second, systematic control input application is performed. The operator applies specific inputs to the accelerator pedal, brake pedal, and steering wheel. Each input is held steady, at various magnitudes, for a predetermined duration. This allows the system to capture stable CAN signals corresponding to each control action.

Third, data recording and signal identification is performed. The system records the CAN signals associated with each control input. The correlation functions, previously developed in [5], [6], are applied to identify which CAN signals are most closely associated with each control input. This step is critical to ensure that the system accurately identifies the control-specific CAN messages, filtering out any unrelated signals.

Finally, GUI-Guided Confirmation is performed. Throughout the calibration, the GUI provides real-time feedback to the operator, visually confirming that the CAN signals are being captured during the process. This reduces the likelihood of errors and helps ensure that the calibration is successfully completed.

Successful completion of the calibration phase ensures that subsequent real-time inference operates on semantically grounded CAN channels during live or simulated vehicle operation. Calibration thus serves as a validation mechanism that supports accurate real-time semantic reverse engineering without requiring prior vehicle knowledge or explicit signal tokenization.

### *3.5 Real-Time Validation through GUI*

A graphical user interface (GUI) was developed to enable interactive visualization and validation of inferred CAN channels during vehicle operation. The GUI allows the operator to observe, select, and verify candidate CAN channels associated with key vehicle controls, including the accelerator pedal, brake pedal, and steering wheel.

During operation, the GUI provides immediate visual feedback as vehicle controls are actuated. For example, steering input is indicated through a steering wheel visualization, while brake and accelerator inputs are displayed through pedal representations. These visualizations update continuously based on the real-time output of the inferred CAN channels, allowing the operator to confirm that observed channel behavior aligns with expected control actions.

The GUI is not used to define or recover CAN signal structure. Instead, it serves as a real-time semantic validation layer, enabling confirmation that the highest-ranked channels produced by the inference pipeline correspond to meaningful vehicle control inputs. This interactive validation step is particularly valuable in real-time scenarios, where rapid confirmation of inference results is required without offline analysis.

Although designed for in-vehicle operation, the GUI was also evaluated using rosbag replay. Data recorded during live vehicle sessions was replayed in a simulated real-time environment, allowing the GUI to operate identically to live use. This ensured consistent validation for both controlled laboratory testing and real-world driving scenarios.

This real-time validation capability provides an essential human-in-the-loop mechanism. It increases confidence in the inferred CAN channel mappings while preserving the system's zero-knowledge and real-time operating assumptions.

Figure 7 shows the GUI during in-vehicle operation, demonstrating the live visualization of vehicle control inputs and inferred CAN channels. Figure 8 presents a functional breakdown of the GUI into its primary components.

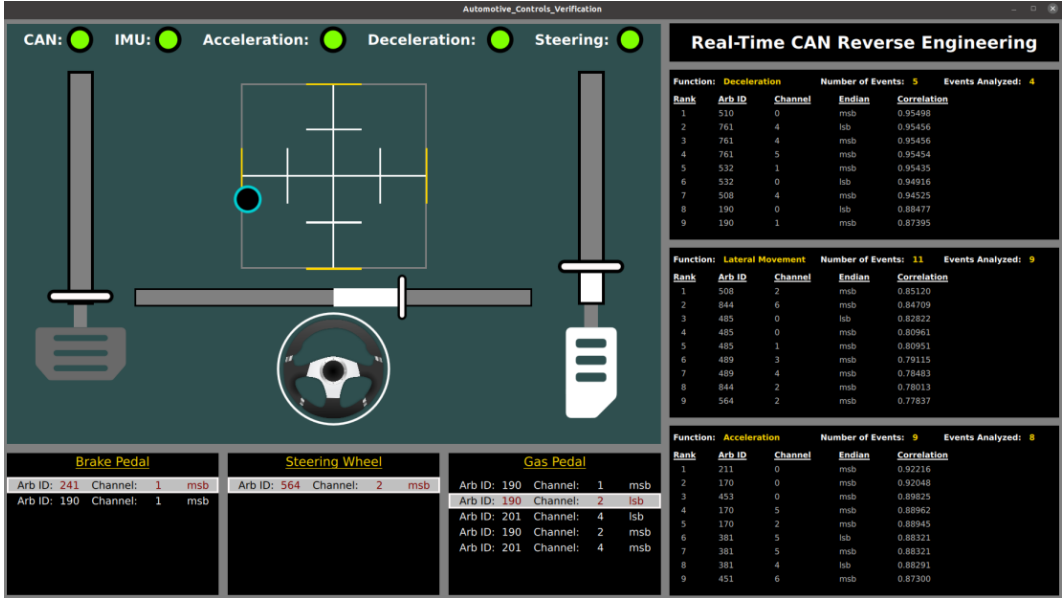

**Figure 7.** Real-time graphical user interface.

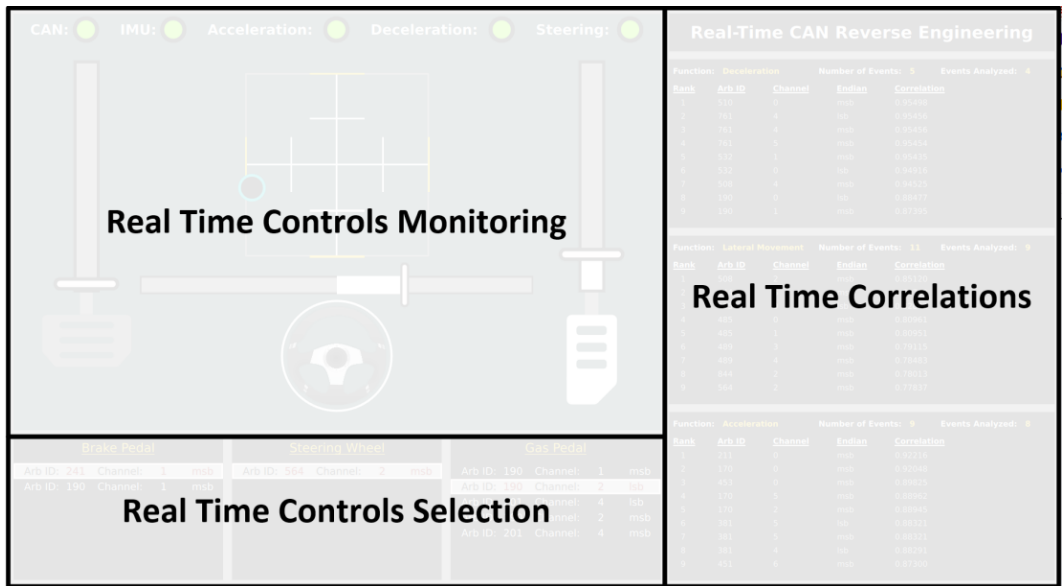

**Figure 8.** Sections within graphical user interface.

## 4. Results

This section presents the results obtained from the real-time application of the proposed CAN reverse engineering methodology. The primary focus of this section is on the main evaluation platform, a 2016 GMC Sierra, which served as the test vehicle for detailed analysis. Results for additional vehicles are provided in Appendix A they demonstrate generalizability across different vehicles and manufacturers.

The primary outcomes of interest are the inferred CAN channels associated with three key vehicle control inputs: the accelerator pedal position, the brake pedal position, and the steering wheel input. Because the proposed approach is designed for real-time operation, system performance metrics – including processing latency and event convergence behavior – are also examined.

To aid in interpretation of the results, event-aligned visualizations are presented for each control category. These plots illustrate the temporal behavior of candidate CAN channels during calibration and driving events, along with the expected physical response which is inferred from inertial measurements. These visualizations serve as qualitative validation aids, allowing the semantic alignment between vehicle actions and inferred CAN channel behavior to be assessed.

All of the results presented in this section were derived from rosbag recordings generated in the prior studies [5], [6]. These recordings captured synchronized CAN, IMU, and GPS data during live vehicle operation. By replaying these datasets in a simulated real-time environment, the proposed system was evaluated under identical conditions to previous work. This ensures direct comparability of results, facilitating the analysis of the impact of the real-time architectural enhancements that have been introduced.

In this section, inferred CAN channels are referenced using a structured naming convention. An example channel identifier is shown in Table 1.

**Table 1.** Identifier 190_msb_0 meaning.

| Field | Value |
|---|---|
| Arbitration ID | 190 |
| Channel Width | 16 bits (uniform hypothesis size used in this study) |
| Channel | 0 |
| Bit Ordering | Most significant bit first (MSB) |

All reported channels use a fixed-width segmented channel hypotheses rather than decoded OEM signal definitions. As discussed in Section 3.3, these channels serve as semantic candidates for inference and validation, not as representations of true CAN signal boundaries.

### 4.1. Accelerator Pedal Results

This section presents the results of real-time semantic inference for accelerator pedal input from a 2016 GMC Sierra 1500. The objective of this analysis is to identify CAN channels whose temporal behavior exhibits strong alignment with accelerator pedal actuation, without relying on prior vehicle knowledge or decoded OEM signal definitions.

Acceleration-related inference was triggered after every three detected acceleration events, as identified by the acceleration event node. A total of fifteen acceleration events were processed in the dataset. At each inference interval, the system ranked candidate CAN channels based on their correlation strength with acceleration-related inertial measurements.

Tables 2 and 3 list the highest-ranked candidate channels that were identified at successive inference stages. Each table lists the arbitration ID, fixed-width channel hypothesis, and correlation value. The correlation values indicate the degree of alignment between candidate

CAN channels and observed acceleration behavior. Higher values indicate stronger temporal correspondence.

All results were obtained using rosbag recordings generated in prior studies [5], [6]. By replaying identical datasets, the evaluation isolates the impact of the proposed real-time architecture while ensuring direct comparability with earlier methods.

Table 2 reports inference results after analyzing three, six, and nine acceleration events. At early stages, multiple candidate channels exhibit comparable correlation values, reflecting limited event context. As additional events are incorporated, correlation rankings begin to stabilize, narrowing the candidate set.

**Table 2.** Accelerator Pedal Position Results Through Nine Events.

| Three Events (37 sec) | | | Six Events (67 sec) | | | Nine Events (97 sec) | | |
|---|---|---|---|---|---|---|---|---|
| ID | Channel | Correlation | ID | Channel | Correlation | ID | Channel | Correlation |
| 201 | msb_4 | 0.6507645 | 190 | msb_2 | 0.483471561 | 201 | msb_4 | 0.54085495 |
| 201 | lsb_4 | 0.6507645 | 201 | msb_4 | 0.483180943 | 201 | lsb_4 | 0.5407765 |
| 190 | msb_1 | 0.6474277 | 201 | lsb_4 | 0.483163946 | 190 | msb_2 | 0.54067352 |
| 190 | lsb_1 | 0.6474277 | 190 | lsb_1 | 0.482545715 | 190 | lsb_1 | 0.53753719 |
| 190 | msb_2 | 0.6474277 | 190 | lsb_2 | 0.466186838 | 190 | lsb_2 | 0.52357252 |
| 190 | lsb_2 | 0.6282368 | | | | | | |

Table 3 presents the results after 12 and 15 acceleration events. At this stage, a clear separation emerges between the highest-ranked and lower-ranked candidates. This convergence behavior shows that semantic inference improves as additional event data becomes available, reinforcing the importance of event accumulation in real-time operation.

**Table 3.** Accelerator Pedal Position Results Through 15 Events.

| 12 Events (168 sec) | | | 15 Events (163 sec) | | |
|---|---|---|---|---|---|
| ID | Channel | Correlation | ID | Channel | Correlation |
| 201 | msb_4 | 0.5664567 | 190 | msb_2 | 0.832201943 |
| 201 | lsb_4 | 0.5663861 | 201 | msb_2 | 0.662868015 |
| 190 | msb_2 | 0.5659167 | 201 | lsb_4 | 0.662665021 |
| 190 | lsb_1 | 0.563388 | 190 | lsb_2 | 0.654124523 |
| 190 | lsb_2 | 0.553342 | 190 | lsb_1 | 0.51528777 |

Figures 9 and 10 provide qualitative validation for two candidate channels that were identified during the accelerator pedal calibration phase. Each figure is divided into four quadrants. The top-left quadrant shows the calibration recording for the selected channel and the top-right quadrant displays the expected accelerator actuation waveform. The bottom-left quadrant illustrates the channel behavior during detected acceleration events and the bottom-right quadrant juxtaposes the candidate channel output with inertial acceleration data from the full recording.

The figures demonstrate strong temporal alignment between the inferred CAN channels and accelerator pedal behavior, despite the use of the fixed-width channel hypotheses, rather than decoded signal tokens. The observed alignment supports the conclusion that semantically meaningful accelerator-related channels can be identified through correlation-based inference alone. It is important to note that these figures were generated specifically for analysis and presentation purposes. Real-time validation of accelerator pedal inference occurs through the graphical user interface, as described in Section 3.5, where an operator can confirm semantic correctness during live or simulated vehicle operations. The figures, thus, serve as illustrative confirmation of the system's real-time inference performance.

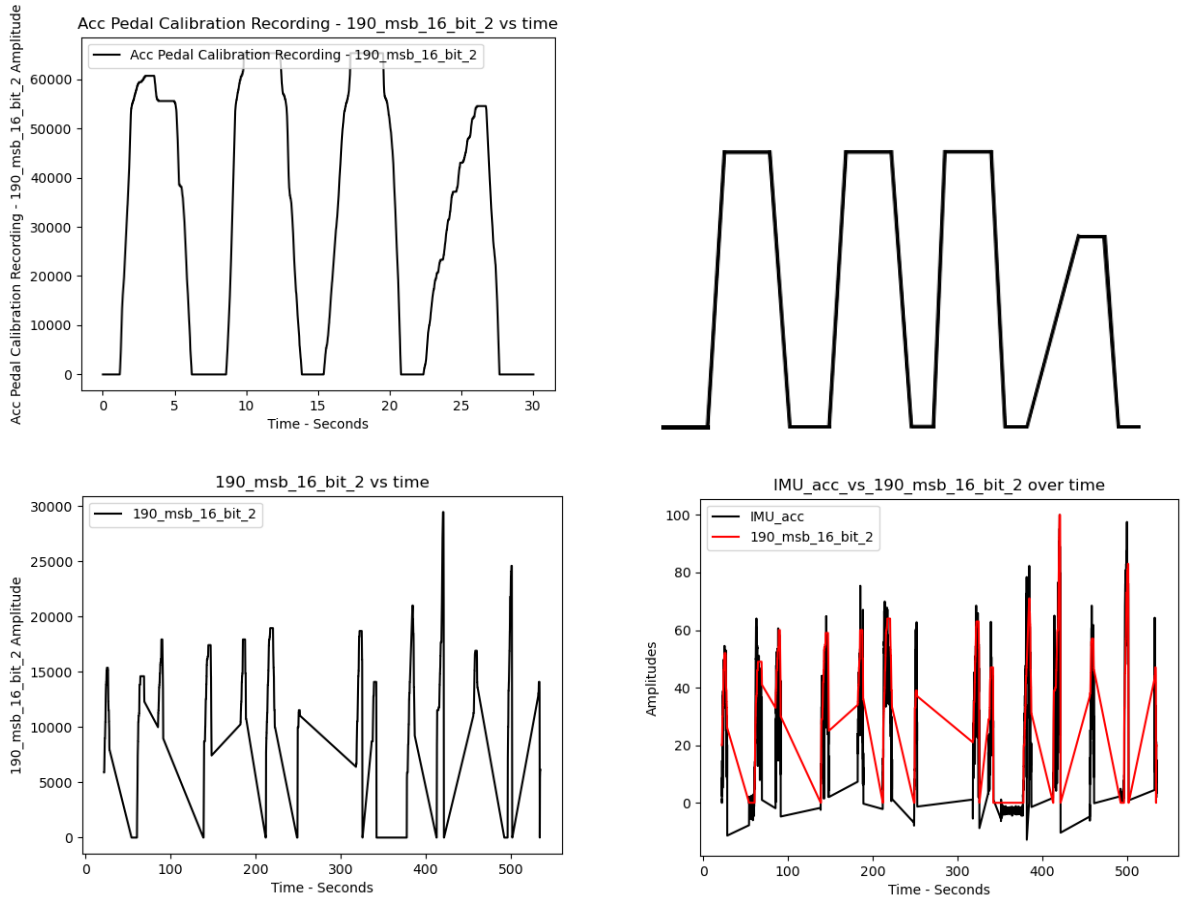

**Figure 9.** 2016 GMC Sierra 1500 accelerator pedal validation one.

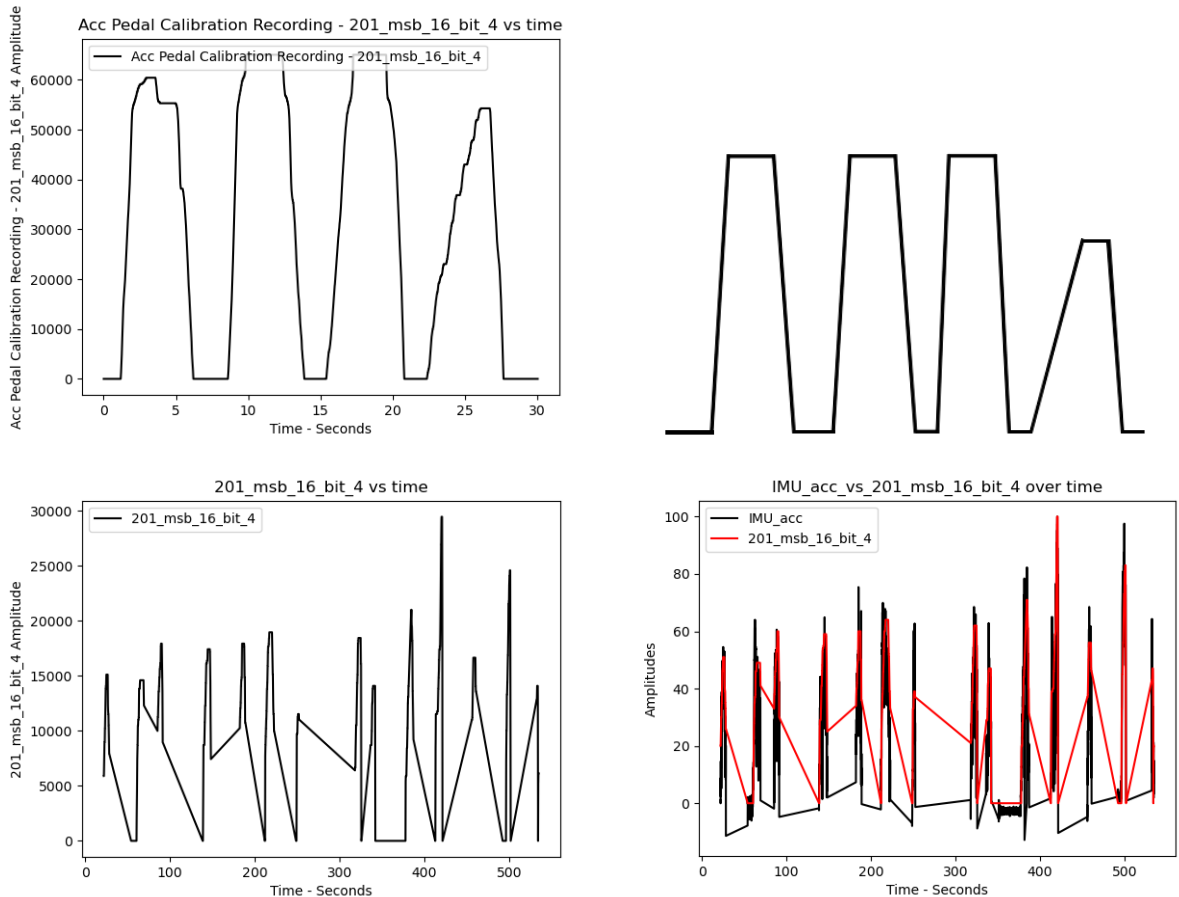

**Figure 10.** 2016 GMC Sierra 1500 accelerator pedal validation two.

## 4.2. Brake Pedal Results

This section presents the results of real-time semantic inference for brake pedal input from the 2016 GMC Sierra 1500. The objective of this analysis is to identify CAN channels whose temporal behavior exhibits strong alignment with vehicle deceleration events, without relying on prior vehicle knowledge or decoded OEM signal definitions.

Deceleration-related inference was triggered after every three newly detected deceleration events, as identified by the deceleration event node. A total of fifteen deceleration events were processed from the dataset. During each inference process, candidate CAN channels were ranked based on their correlation strength with deceleration-related inertial measurements.

Tables 4 and 5 summarize the highest-ranked candidate channels identified at successive inference stages. Each table reports the arbitration ID, fixed-width channel hypothesis, and associated correlation value. These correlation values represent the degree of alignment between the candidate CAN channels and the observed deceleration behavior.

All of these results were obtained using rosbag recordings generated in prior studies [5], [6]. By replaying identical datasets, the evaluation ensures direct comparability with earlier work and isolates the effects of the proposed real-time architectural enhancements.

Table 4 presents inference results after analyzing three, six, and nine deceleration events. Even at early stages, a small set of candidate channels are consistently identified with relatively strong correlation values. This reflects the pronounced and repeatable nature of brake pedal actuation, as compared to other vehicle control inputs.

**Table 4.** Brake pedal position results through nine events.

| Three Events (35 sec) | | | Six Events (78 sec) | | | Nine Events (87 sec) | | |
|---|---|---|---|---|---|---|---|---|
| ID | Channel | Correlation | ID | Channel | Correlation | ID | Channel | Correlation |
| 190 | msb_1 | 0.7033459 | 190 | msb_1 | 0.734831991 | 190 | msb_1 | 0.72287794 |
| 241 | msb_1 | 0.693279 | 241 | msb_1 | 0.725404702 | 241 | msb_1 | 0.71689376 |

Table 5 presents the results after 12 and 15 deceleration events. Note that the leading candidate channels were dominant across all inference intervals. This indicates that brake-related semantic inference is robust and requires fewer events to reach high confidence, relative to other control inputs.

**Table 5.** Brake pedal position results through 15 events.

| 12 Events (132 sec) | | | 15 Events (217 sec) | | |
|---|---|---|---|---|---|
| ID | Channel | Correlation | ID | Channel | Correlation |
| 190 | msb_1 | 0.7187651 | 190 | msb_1 | 0.744118782 |
| 241 | msb_1 | 0.7151189 | 241 | msb_1 | 0.742519918 |

Figures 11 and 12 provide qualitative validation for the two candidate channels identified during the brake pedal calibration phase. Each figure follows the same four-quadrant format used for the accelerator pedal analysis. The top-left quadrant shows the calibration recording for the selected channel, and the top-right quadrant displays the expected brake pedal actuation waveform. The bottom-left quadrant illustrates channel behavior during detected deceleration events, and the bottom-right quadrant juxtaposes candidate channel output and inertial deceleration data.

These visualizations demonstrate strong temporal alignment between inferred CAN channels and brake pedal behavior, despite the use of the fixed-width channel hypotheses, rather than decoded signal tokens. The observed alignment supports the conclusion that semantically meaningful brake-related channels can be identified through correlation-based inference alone.

As with the accelerator pedal results, these figures were generated specifically for analysis and presentation. Real-time validation of brake pedal inference occurs through the graphical user interface, as described in Section 3.5, where the operator can confirm semantic correctness during live or simulated vehicle operation. The figures serve to confirm real-time inference performance.

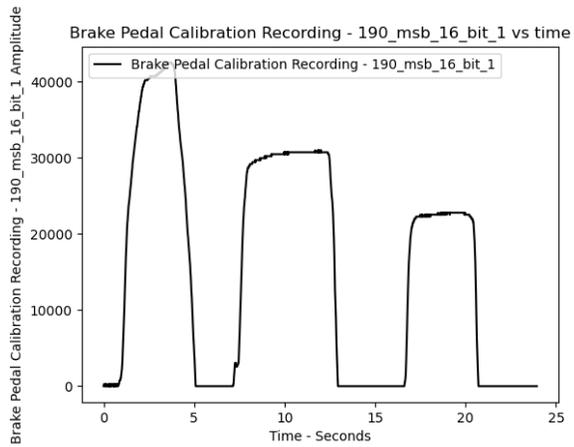
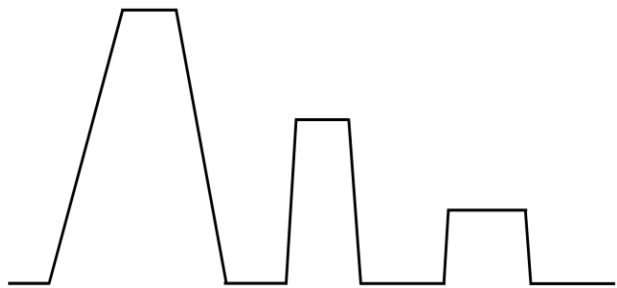
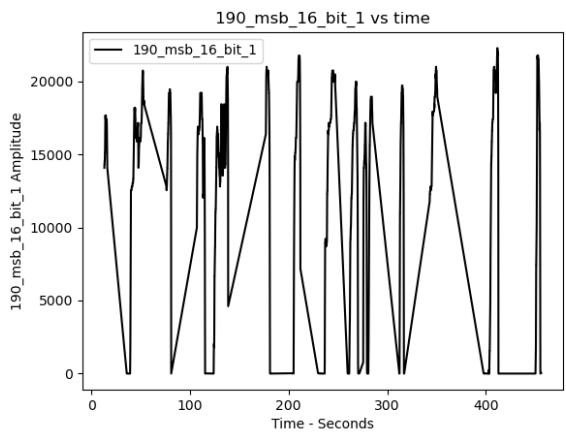
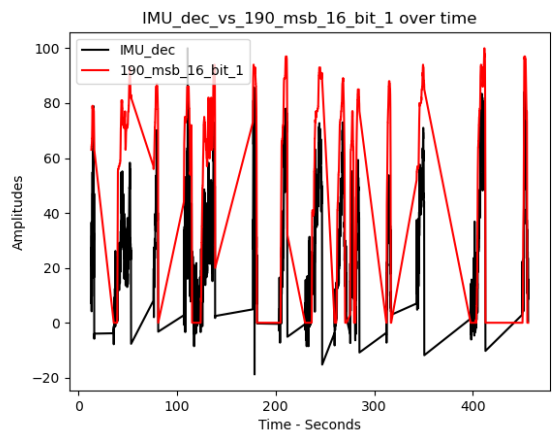

**Figure 11.** 2016 GMC Sierra 1500 Brake pedal validation one.

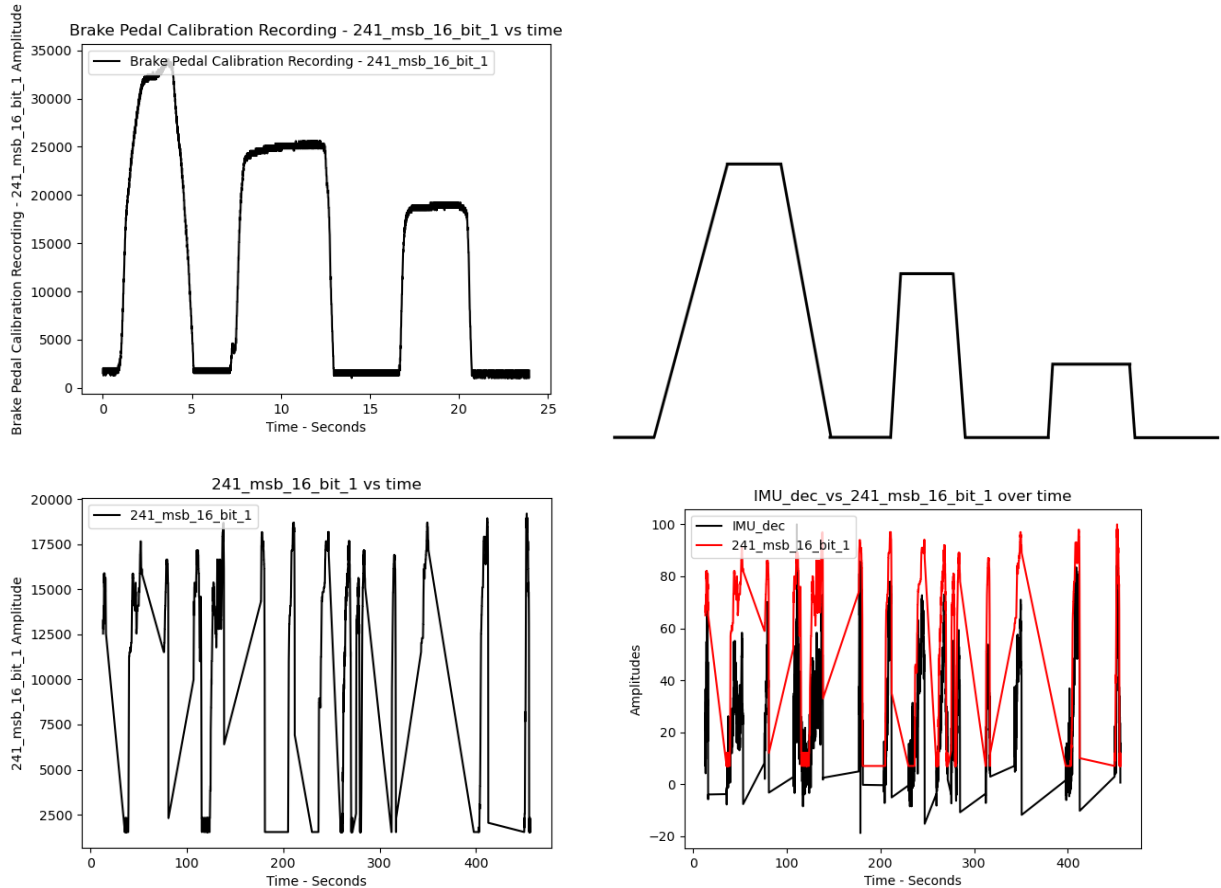

**Figure 12.** 2016 GMC Sierra 1500 Brake pedal validation two.

*4.3. Steering Wheel Results*

This section presents the results of real-time semantic inference for steering wheel input from the 2016 GMC Sierra 1500. The objective of this analysis is to identify CAN channels whose temporal behavior exhibits strong alignment with steering-induced lateral motion, without relying on prior vehicle knowledge or decoded OEM signal definitions.

Steering-related inference was triggered after every three detected steering events, as identified by the steering event node. A total of fifteen steering events were processed from the dataset. At each inference interval, candidate CAN channels were ranked based on their correlation strength with the inertial measurements associated with the lateral acceleration and angular motion.

The results presented in this subsection were obtained using rosbag recordings generated in prior studies [5], [6]. By replaying these datasets, the evaluation maintains direct comparability with the earlier work, demonstrating the impact of the proposed real-time, event-driven inference architecture.

Table 6 summarizes the highest-ranked candidate channels identified after analyzing 3 and 6 steering events. At the earliest inference stages, multiple candidate channels exhibit moderate correlation values, reflecting the inherently noisier and more continuous nature of steering input, as compared to pedal-based controls. Unlike braking events, which tend to produce discrete and repeatable actuation patterns, steering behavior can include smaller corrective motions that reduce early-stage correlation.

**Table 6.** Steering wheel position results through six events.

| Three Events (44 sec) | | | Six Events (70 sec) | | |
|---|---|---|---|---|---|
| ID | Channel | Correlation | ID | Channel | Correlation |
| 801 | msb_0 | 0.5567214 | 564 | msb_2 | 0.557990515 |
| 564 | msb_2 | 0.5358519 | | | |
| 1020 | msb_3 | 0.500205 | | | |
| 1020 | msb_6 | 0.500205 | | | |
| 1017 | msb_1 | 0.4684405 | | | |

Table 7 presents the results after analyzing nine and 12 steering events. Like with the six event data, a single candidate channel is consistently the dominant hypothesis. This stabilization indicates that cumulative event aggregation improves semantic separation between steering-related CAN activity and unrelated background traffic.

**Table 7.** Steering wheel position results through 12 events.

| Nine Events (118 sec) | | | 12 Events (177 sec) | | |
|---|---|---|---|---|---|
| ID | Channel | Correlation | ID | Channel | Correlation |
| 564 | msb_2 | 0.58411429 | 564 | msb_2 | 0.573139 |

Finally, Table 8 presents the results after all fifteen steering events have been analyzed. The single CAN channel maintains the highest correlation value, which has notably increased across the inference intervals. This indicates strong and repeatable semantic alignment with steering input.

**Table 8.** Steering Wheel Position Results Through 15 Events.

| 15 Events (382 sec) | | |
|---|---|---|
| ID | Channel | Correlation |
| 564 | msb_2 | 0.608047374 |

Figure 13 provides qualitative validation of the identified steering-related CAN channel using data recorded during the steering wheel calibration phase. The figure follows the same four-quadrant format used for the accelerator and brake pedal analyses. The top-left quadrant displays the calibration recording for the selected channel, and the top-right quadrant shows the expected steering actuation waveform. The bottom-left quadrant illustrates channel behavior during the detected steering events, and the bottom-right quadrant juxtaposes the candidate channel output and the inertial steering-related measurements for the full recording duration.

The observed temporal alignment demonstrates that semantically meaningful steering-related CAN channels can be identified using the fixed-width channel hypotheses, even in the absence of explicit signal tokenization or decoded message definitions. As with the other control inputs, this result shows the effectiveness of correlation-based semantic inference for real-time reverse engineering.

As with the other controls, the figures presented here were generated for analysis and presentation purposes. The real-time validation of steering inference is performed through the GUI, as described in Section 3.5, where the operator can visually confirm semantic correctness during live or simulated vehicle operation. The figures serve as illustrative confirmation of real-time inference performance.

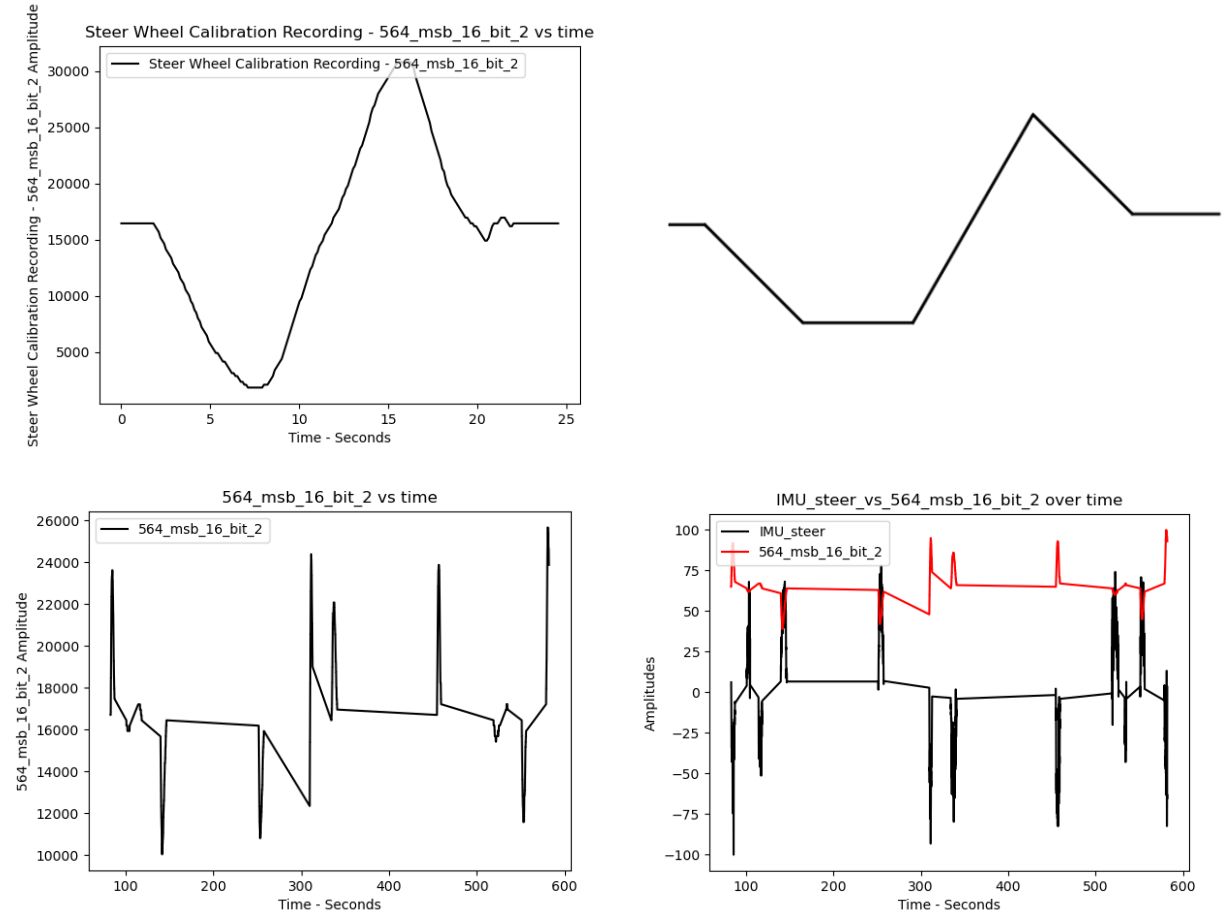

**Figure 13.** 2016 GMC Sierra 1500 steering wheel validation.

## 5. Discussion

This section interprets the experimental results obtained from the proposed real-time CAN reverse-engineering system. The discussion focuses on the importance of the observed inference behavior, the validation and accuracy characteristics, and the assessment of validity through independent comparison. The impact of architectural design choices are also discussed, as are the limitations and implications for future research.

### *5.1. Overview of the Findings*

The primary objective of this work was to determine whether the CAN channels associated with vehicle control inputs can be identified in real time without prior knowledge of the vehicle's CAN architecture. The system consistently inferred CAN channels corresponding to accelerator pedal input, brake pedal input, and steering wheel motion, This was demonstrated for the 2016 GMC Sierra 1500, as presented in Section 4, as well as for additional vehicles, presented in Appendix A.

One exception to this was observed for the steering wheel inference on the 2009 Chevy Impala. This outcome is likely due to the architectural differences in vehicle networks, such as steering-related signals residing on a separate bus or being absent from the high-speed CAN interface accessible during testing. This behavior is consistent with variability in CAN architectures across manufacturers and model years rather than a deficiency in the inference method.

These results indicate that the proposed approach generalizes across multiple vehicle platforms without reliance on vehicle-specific assumptions. The method does not require prior signal definitions, arbitration ID knowledge, or manufacturer documentation. This directly addresses a key limitation of existing reverse-engineering workflows.

Evaluation was performed using rosbag recordings collected in prior studies [5], [6]. Replaying identical datasets eliminated variability associated with live data collection and enabled direct comparison with the previous offline approaches. This controlled evaluation isolates the effect of the real-time architecture and confirms that the observed performance differences are due to methodological changes rather than differences in data.

Inference behavior across successive control events revealed a clear convergence pattern. Channel identification based on the first three detected events was occasionally unstable, whereas inference reliability improved with more identification events. For example, accelerator pedal inference on the 2016 GMC Sierra 1500 stabilized after six events. This behavior indicates that the method benefits from a modest accumulation of event context and suggests that further optimization could potentially reduce convergence time in real-time deployments.

*5.2. Validation and Accuracy of Identified CAN Channels*

A calibration phase was used to constrain inference to signals associated with driver control inputs. As described in Section 3.4, calibration is performed under controlled vehicle conditions. With the engine off for accelerator and brake calibration and the engine on for steering calibration for steering power assistance, while the operator follows GUI-guided prompts to apply pedal and steering inputs. During this period, each control is actuated to its safe minimum and maximum positions, allowing the system to observe the corresponding signal range and establish a control-specific response profile. The highest responding channels identified during this process are retained as a constrained candidate set for subsequent inference and GUI-based validation. Because the system has already established the expected response range for each control, real-time validation can interpret relative input magnitude while limiting the influence of unrelated CAN traffic.

Validation was performed by comparing the inferred channel behavior against the expected control waveforms, as shown in Figures 9 to 13. These figures were generated for this analysis and presentation and were not part of the real-time inference loop. Real-time validation instead occurred through the graphical user interface, which allowed an operator to actuate vehicle controls and immediately observe corresponding inferred signal behavior. This provided direct confirmation of semantic correctness during live and simulated operations.

Reusing the rosbag datasets from the prior studies [5], [6], which contain the same synchronized CAN and IMU recordings, also allowed validation through comparison to this prior work. Using the same recordings, the near real-time pipeline converged on the same dominant control-related channel hypotheses (see Section 4 and Appendix A) as in [5, 6]. The results showed that transitioning to real-time execution does not degrade inference accuracy, relative to offline analysis.

Correlation strength generally increased as additional control events were processed, as shown in Tables 6 through 8 for steering, acceleration, and braking, respectively. Inference confidence improving over time is well suited to continuous operation scenarios, such as autonomous systems and real-time diagnostics.

Displacement is another important characteristic of the inference process. As shown in Tables 6 to 8, an initially identified steering-related channel can be displaced as additional events are observed. This behavior shows that the system refines its hypotheses dynamically and is capable of rejecting early inaccurate identifications as more evidence becomes available. This adaptive refinement directly contributes to inference robustness.

Notably, semantic alignment with known actuator functions emerged, despite the use of fixed-width channel hypotheses rather than true OEM signal tokens. This suggests that control inference does not require precise signal boundaries and that the approach is tolerant of approximate segmentation.

### 5.3. External Validity

External validity was assessed through comparison with comma.ai's openDBC repository [25], which provides independently reverse-engineered CAN message definitions. The proposed method does not use the DBC information during inference, and no manufacturer-specific signal definitions were available to the system during processing.

After inference was completed, the highest-ranked channels that were associated with accelerator input, braking input, and steering angle were compared, at the message level, with the corresponding openDBC definitions for the evaluated platform [25].

For example, for the 2016 GMC Sierra 1500, the system consistently identified arbitration ID 241 as the dominant brake-related message during deceleration events, as shown in Tables 4 and 5. The GM Global A DBC file indicates that message 241 contains the signal *BrakePedalPosition*, defined as an 8-bit field beginning at bit 15 of the payload. Figure 14 shows the relevant excerpt from the openDBC definition for message 241 (EBCMBrakePedalPosition) and the *BrakePedalPosition* signal defined at bit 15 with length 8 bits.

```
85      BO_ 241 EBCMBrakePedalPosition: 6 K17_EBCM
86        SG_ BrakePressed : 1|1@0+ (1,0) [0|1] "" XXX
87        SG_ BrakePedalPosition : 15|8@0+ (1,0) [0|255] ""  NEO
```

**Figure 14.** Excerpt from comma.ai openDBC (GM Global A, message 241) [25].

The fixed-width 16-bit channel hypothesis, that was chosen for this work, encompassed this bit region. Although precise OEM tokenization was not performed, the inferred channel overlapped with the true encoded brake position field within the message payload. This region corresponds to the 16-bit channel segment, shown in Figure 11, where strong temporal alignment with brake actuation was observed.

While the proposed method does not determine the exact bit-level tokenization, the observed agreement with established DBC definitions indicates that semantic alignment with vehicle dynamics allows the identification of the location of the true control encodings within the CAN payload.

### 5.4. Method Enhancements and Their Impact

The observed real-time performance results primarily from architectural design choices rather than changes to the underlying inference algorithm. The adoption of the event-based architecture, discussed in Section 3.2.1, constrained correlation analysis to discrete acceleration, deceleration, and steering actions. By performing inference only within bounded event windows, the system avoided continuous stream-wide processing and instead concentrated computation on intervals containing meaningful control dynamics. This reduced the data volume per inference cycle and improved the signal-to-noise conditions under which correlation was evaluated.

In addition to temporal restriction, the system architecture also separated event detection, correlation analysis, filtering, and visualization into independent processing threads within the ROS framework. This multithreaded design reduced blocking between the acquisition and analysis tasks, allowing inference to proceed concurrently with data collection and user interaction. The parallel execution model supported near real-time responsiveness, while maintaining processing behavior.

Together, these architectural choices enabled inference under real-time constraints without altering the underlying correlation methodology. The improvements observed in convergence behavior and processing latency, therefore, reflect design enhancements rather than changes to the inference strategy.

### 5.5. Limitations and Challenges

The calibration process, while aiding inference, introduced several practical constraints. Brake pedal calibration was affected by the pressure buildup after initial brake application, and steering calibration required the engine to be running to engage power steering. Reducing or automating the calibration requirements represents a clear opportunity for future improvement.

Broader evaluation across additional vehicle makes, models, and model years would allow further analysis of the proposed technique's generalizability. Although performance was consistent, under the controlled conditions, expanded testing would allow the characterization of behavior across a wider range of CAN architectures.

Real-time operations also introduced challenges related to operator involvement during validation. While performance remained robust, reducing the need for manual confirmation, through increased automation, could further improve reliability and system usability.

Live driving tests produced behavior consistent with rosbag-based evaluation. The quantitative results from these tests are not included. Future work could also incorporate and analyze live operational data.

### 5.6. Broader Implications and Future Work

Real-time CAN reverse engineering, without prior vehicle knowledge, has direct implications for aftermarket autonomy, diagnostics, and cybersecurity. The ability to infer control-related signals dynamically enables platform-agnostic system integration without OEM documentation.

The approach is also relevant to automotive cybersecurity. Real-time identification of control channels enables monitoring for anomalous or unauthorized CAN activity and provides a foundation for intrusion detection and response mechanisms. The generalizability of the proposed approach facilitates the creation of intrusion detection systems that can operate across multiple manufacturers' vehicles.

Although the method successfully identifies control-related signals, inference alone is not sufficient for closed-loop vehicle control. Future work will be needed to examine the contextual dependencies and system integration requirements necessary for safe actuation.

Integrating real-time signal tokenization is a natural extension of this work. While semantic inference proved effective, combining inference with precise signal delineation could further improve accuracy and enable full end-to-end CAN reverse engineering in real-time systems.

## 6. Conclusions and Future Work

The work presented herein demonstrated that CAN channels associated with primary vehicle control inputs – specifically accelerator pedal position, brake pedal position, and steering wheel motion – can be inferred in real time and without prior knowledge of a vehicle's internal CAN architecture. By combining inertial measurements, passive CAN traffic monitoring, and an event driven software architecture, the proposed system is able to autonomously identify the control related CAN channels during live or replayed vehicle operations.

Experimental results from multiple vehicles show that the real time system consistently identifies the same control related channels that were previously recovered using offline reverse engineering techniques. These results were achieved while reducing the processing latency and eliminating the need for additional hardware, such as GPS. This demonstrated that real time semantic inference can match offline accuracy with substantially lower system complexity.

The proposed architecture was intentionally structured to support extensibility. Its modular, event driven design and reliance on abstracted signal representations allows the framework to accommodate additional vehicle functions and sensing modalities without requiring vehicle specific CAN knowledge. Because the approach does not depend on predefined signal definitions or manufacturer specific CAN standards, it remains robust to variation across platforms and adaptable to alternative in vehicle communication technologies beyond CAN.

While this study establishes the feasibility of real time, vehicle agnostic CAN reverse engineering using semantic inference alone, several areas for extension remain. Integrating real time CAN signal tokenization would enable structural signal recovery, alongside semantic identification. This could potentially provide a complete CAN reverse engineering solution. In addition, evaluation with more vehicles and driving scenarios would allow additional analysis and validation of the proposed approach.

This work has shown that meaningful vehicle control semantics can be extracted from CAN traffic in real time without prior system knowledge. The results support the viability of semantic inference as a foundation for applications in aftermarket autonomy, vehicle diagnostics, and automotive cybersecurity, where rapid deployment and platform independence are critical.

**Appendix A**

This appendix provides the results from testing additional vehicles beyond the primary 2016 GMC Sierra 1500. The vehicles tested include a 2021 GMC Sierra 2500, a 2022 Chevrolet Traverse, a 2006 Volvo XC70, a 2009 Chevrolet Impala, and a 2016 Ford Fusion. For each vehicle, tables summarize the discovered CAN channels related to the accelerator pedal, brake pedal, and steering wheel. Thes are followed by figures that present the correlation of the identified CAN channels during calibration.

Each vehicle's data is structured as follows: Tables present the correlation values for the identified CAN channels for three, six, nine, and – sometimes – twelve events. They illustrate improved accuracy as more events are processed.

Figures display CAN channel data over time, providing a visual comparison between the observed CAN signals and the expected control inputs. Each figure is divided into four quadrants. The top-left quadrant shows data recorded during the calibration process for the selected CAN channel. The top-right quadrant illustrates the expected waveform that mirrors the vehicle control (e.g., accelerator, brake, steering) being actuated.

The bottom-left quadrant presents the recorded values of the CAN channel over the identified events (e.g., acceleration, braking). Finally, the bottom-right quadrant typically overlays the CAN channel data with corresponding IMU data (e.g., acceleration, deceleration) for validation. In cases where there is a significant magnitude difference, leading to disproportionate scaling, only the IMU data is displayed to ensure clarity and accuracy.

These figures show that the identified channels correlate well with the physical actions of the vehicle's controls. This demonstrates the methodology's robustness.

*A.1. 2021 GMC Sierra 2500 Results*

This section presents the results for the 2021 GMC Sierra 2500, following the structure outlined at the beginning of this appendix. The tables and figures provided here depict the identified CAN channels related to the accelerator pedal, brake pedal, and steering wheel. The results for this vehicle were consistent with those observed for the 2016 GMC Sierra 1500, demonstrating the system's adaptability to different model years within the same vehicle line.

Table A1 summarizes the correlation values for the CAN channels associated with the accelerator pedal across three and six events. Although absolute correlation values decreased in this instance, the dominant channel candidates remained consistent. The stability in ranking suggests that the proposed method converges, in terms of channel identification, even when the correlation magnitude varies.

**Table A1.** Acceleration results.

| | Three Events | | | Six Events | |
|---|---|---|---|---|---|
| ID | Channel | Correlation | ID | Channel | Correlation |
| 201 | msb_4 | 0.592652 | 190 | msb_2 | 0.455850662 |
| 401 | lsb_6 | 0.5924073 | 190 | lsb_2 | 0.455794907 |
| 401 | msb_6 | 0.5924073 | 201 | msb_4 | 0.45566142 |
| 190 | msb_2 | 0.59167 | 401 | lsb_6 | 0.455429259 |
| 190 | lsb_2 | 0.5813739 | 401 | msb_6 | 0.455429259 |
| 190 | msb_1 | 0.5184387 | 190 | msb_1 | 0.408723258 |

Figure A1 visualizes the correlation results for the first identified CAN channel, related to the accelerator pedal. This figure allows for direct comparison between the CAN channel data and the expected acceleration waveform, helping to validate the accuracy of the system in identifying the correct channels.

It is noteworthy that, unlike other vehicles tested, the accelerator pedal calibration recording for the 2021 GMC Sierra 2500 appears to have ended prematurely. This is indicated by the shorter duration of the recording and the fewer instances of pedal depressions and releases compared to other vehicles, like the 2016 GMC Sierra 1500. Nonetheless, the system successfully identified the relevant CAN channels, as evidenced by the correlation values.

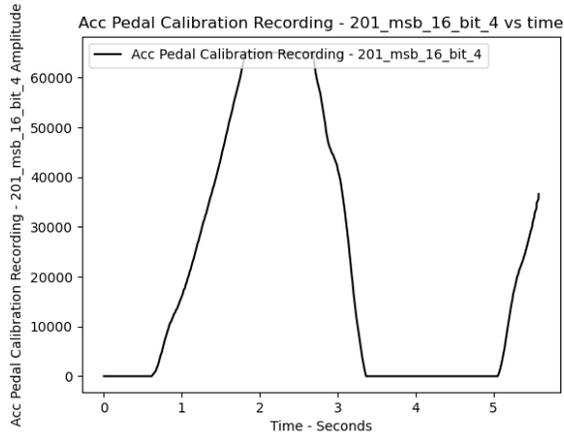
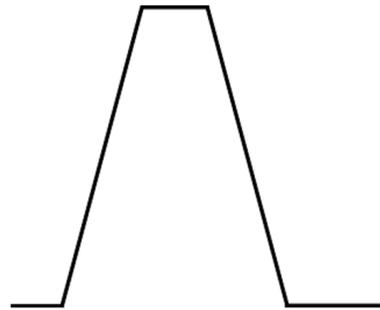
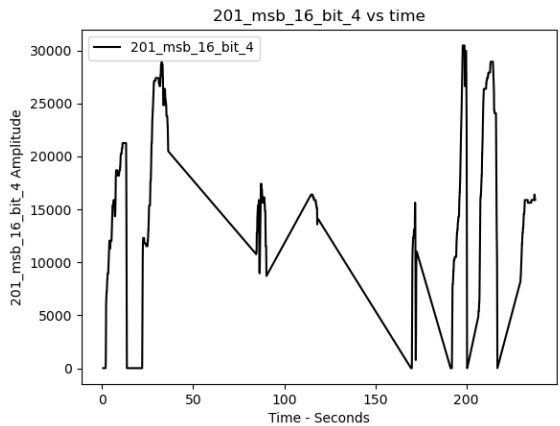
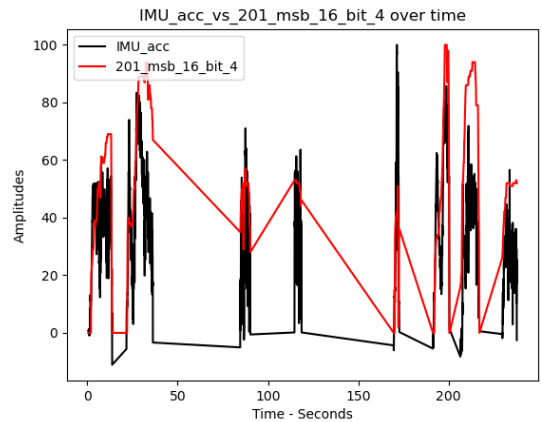

**Figure A1.** 2021 GMC Sierra 2500 accelerator pedal validation one.

Figure A2 provides a similar visualization for a second identified CAN channel that is also related to the accelerator pedal. The comparison between the recorded CAN data and the expected acceleration patterns further supports the system's effectiveness in accurately identifying relevant channels.

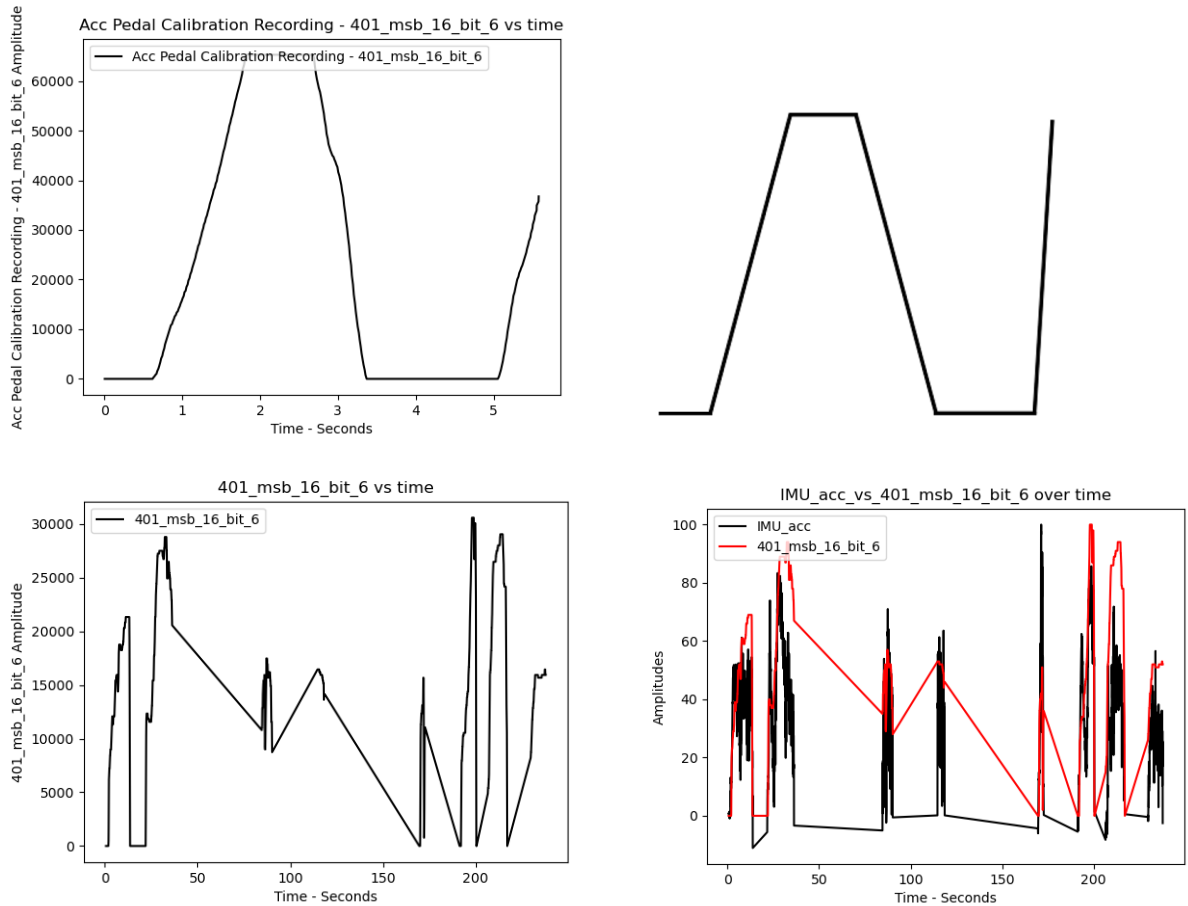

**Figure A2.** 2021 GMC Sierra 2500 accelerator pedal validation two.

Table A2 presents the results for the brake pedal, showing the identified channels and their correlation values for three and six events. The data illustrates how analyzing additional events contributes to refining the identification of the correct CAN channels related to brake pedal input. This is indicated by the reduced set of identified channels in Table A3. This refinement is crucial in improving the system's accuracy.

**Table A2.** Deceleration results.

| Three Events | | | Six Events | | |
| --- | --- | --- | --- | --- | --- |
| ID | Channel | Correlation | ID | Channel | Correlation |
| 190 | msb_1 | 0.8413931 | 190 | msb_1 | 0.820976929 |
| 241 | msb_3 | 0.8410353 | 241 | msb_3 | 0.820216205 |
| 241 | msb_1 | 0.8410352 | 241 | msb_1 | 0.820215724 |
| 241 | msb_4 | 0.8410346 | 241 | msb_4 | 0.820214871 |
| 241 | lsb_1 | 0.8409994 | 241 | lsb_1 | 0.820144794 |
| 1236 | msb_6 | 0.5367574 | 1236 | msb_0 | 0.471968615 |
| 1236 | msb_0 | 0.5366313 | 1236 | msb_6 | 0.471928241 |
| 1236 | msb_2 | 0.5364371 | 1236 | msb_2 | 0.471881776 |
| 1236 | msb_4 | 0.5362001 | 1236 | msb_4 | 0.471576972 |

**Table A3.** Deceleration results.

| | Nine Events | |
|---|---|---|
| ID | Channel | Correlation |
| 190 | msb_1 | 0.78243338 |
| 241 | msb_4 | 0.78230569 |
| 241 | msb_3 | 0.78230525 |
| 241 | msb_1 | 0.78230501 |
| 241 | lsb_1 | 0.78227625 |

Figure A3 presents the time series data for the first identified brake pedal CAN channel for the 2021 GMC Sierra 2500. It shows how the channel's recorded data aligns with the projected brake pedal movements. The figure demonstrates the system's ability to capture and analyze brake pedal input in real time.

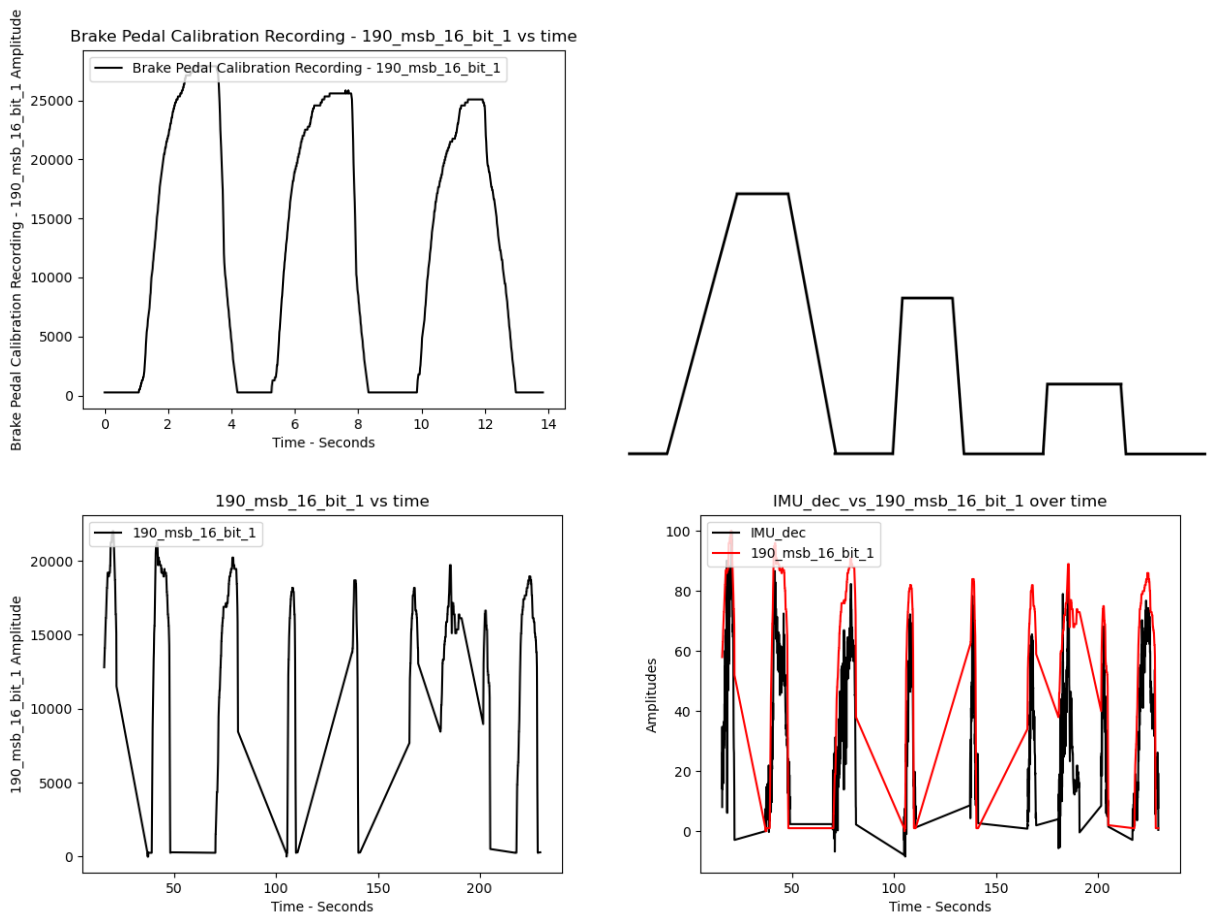

**Figure A3.** 2021 GMC Sierra 2500 brake pedal validation one.

Figure A4 provides a similar time series visualization for the second identified brake pedal CAN channel. The consistent correlation between the recorded CAN data and the expected brake pedal patterns demonstrates the system's ability to accurately identify relevant channels.

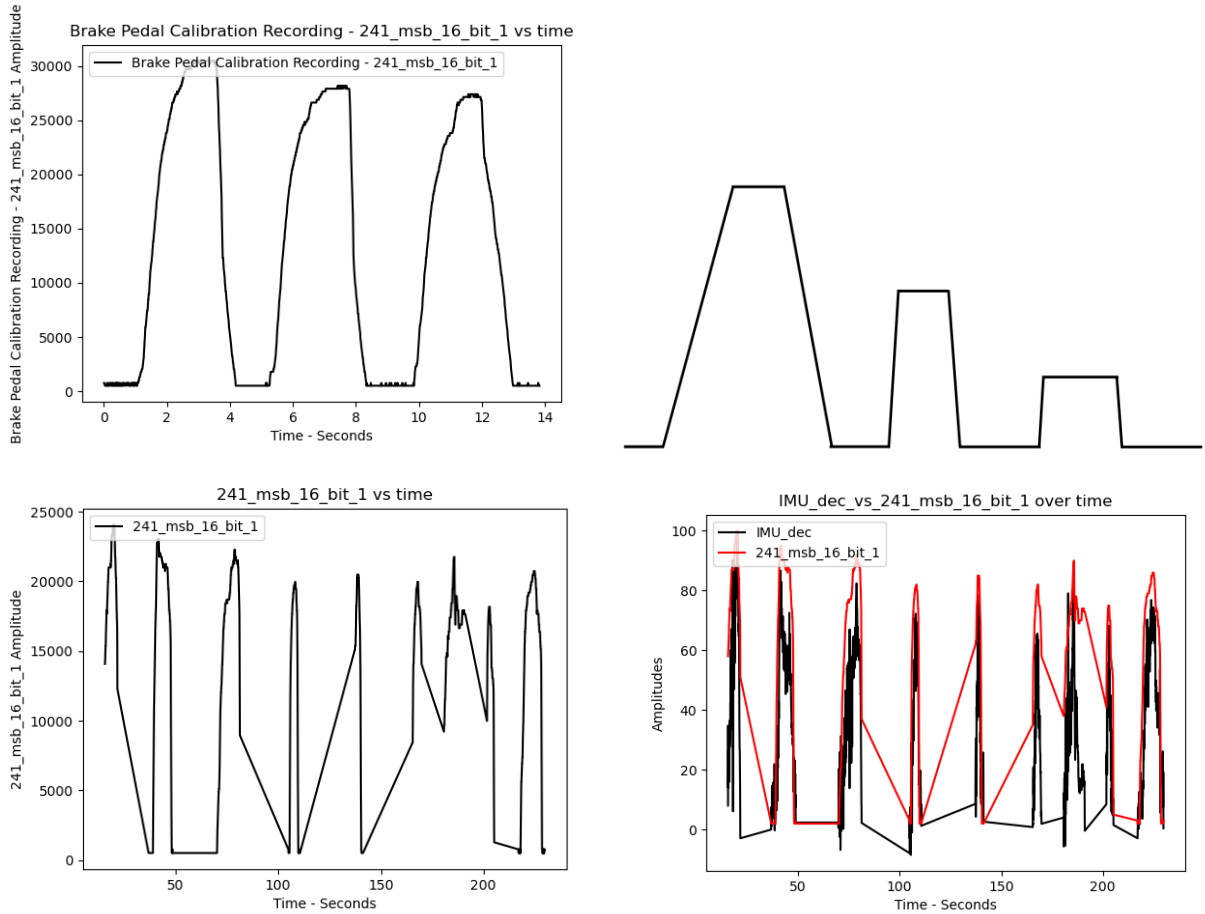

**Figure A4.** 2021 GMC Sierra 2500 brake pedal validation two.

Table A4 presents the results for the steering wheel, showing the correlation values for three and six events. The data reveals a high correlation for the identified CAN channel associated with steering inputs, demonstrating the system's capability to accurately identify relevant channels with fewer events.

**Table A4.** Steering results.

|  | Three Events |  |  | Six Events |  |
|---|---|---|---|---|---|
| ID | Channel | Correlation | ID | Channel | Correlation |
| 564 | msb_2 | 0.883453835 | 564 | msb_2 | 0.8323168 |

Table A5 provides the correlation value for the steering wheel CAN channel from nine events. The high correlation value shown in this table confirms the system's reliability in identifying CAN channels associated with steering movements over multiple events.

**Table A5.** Steering results.

|     | Nine Events |             |
| --- | ----------- | ----------- |
| ID  | Channel     | Correlation |
| 564 | msb_2       | 0.86088381  |

Figure A5 displays the time series data for the steering wheel CAN channel, illustrating how the channel's data corresponds to the expected steering wheel movements. This figure demonstrates the accuracy of the system in capturing and analyzing steering inputs in real time.

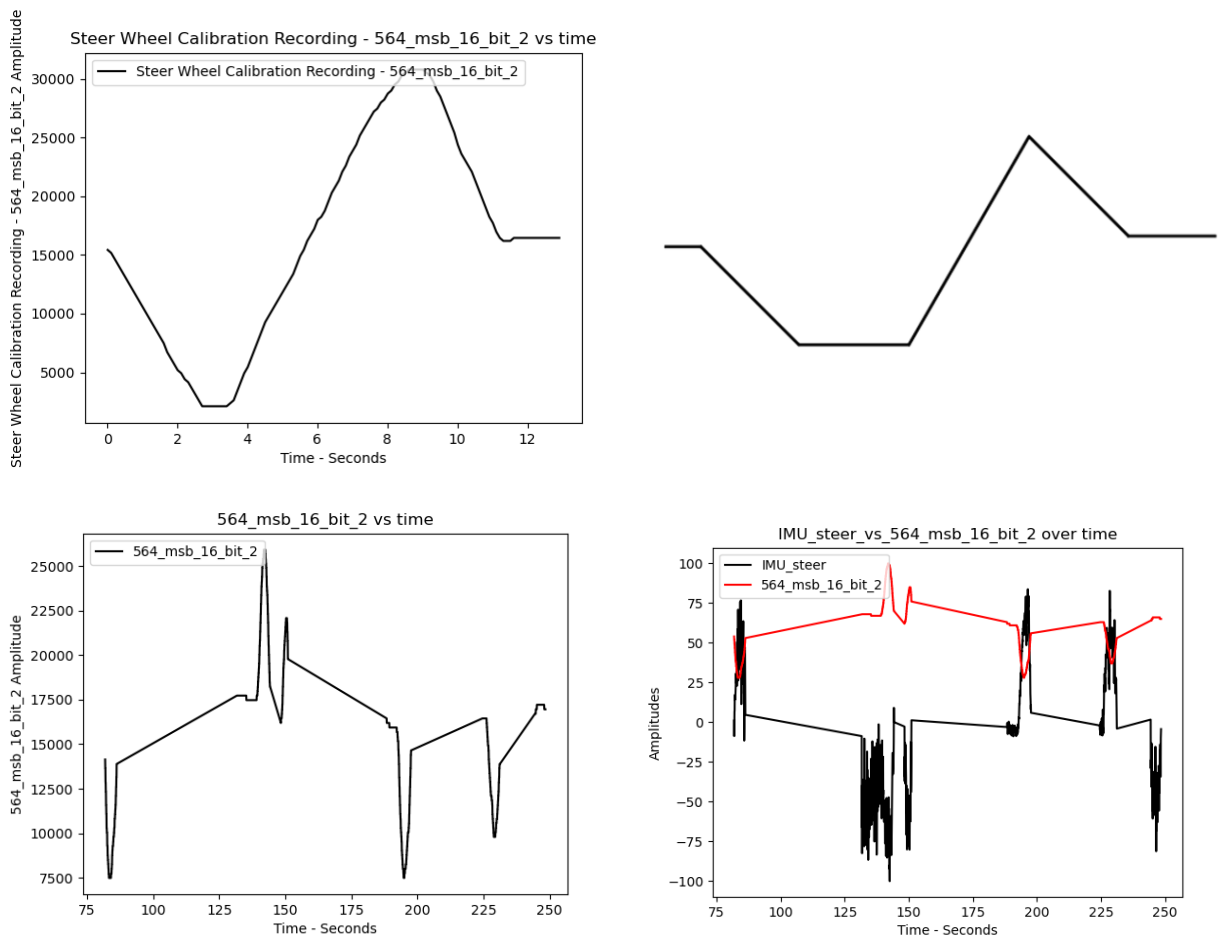

**Figure A5.** 2021 GMC Sierra 2500 steering wheel validation.

*A.2. 2022 Chevrolet Traverse Results*

The 2022 Chevrolet Traverse results also demonstrate the system's capability to reverse engineer CAN signals in real-time across different vehicle makes and models. The consistency in identifying CAN channels across multiple events suggests that the method is effective across the General Motors family of vehicles.

Table A6 presents the correlation values for the CAN channels associated with the accelerator pedal across three and six identification events. The increasing correlation values from the additional events demonstrates the system's robustness in identifying the relevant CAN channels.

**Table A6.** Acceleration results.

| Three Events | | | Six Events | | |
|---|---|---|---|---|---|
| ID | Channel | Correlation | ID | Channel | Correlation |
| 190 | msb_2 | 0.762183507 | 190 | msb_2 | 0.77136402 |
| 401 | msb_6 | 0.759979326 | 401 | lsb_6 | 0.77016616 |
| 401 | lsb_6 | 0.759979326 | 401 | msb_6 | 0.77016616 |
| 201 | msb_4 | 0.759978372 | 201 | msb_4 | 0.77010666 |
| 201 | lsb_4 | 0.759131161 | 201 | lsb_4 | 0.76939043 |
| 451 | lsb_6 | 0.756636335 | 451 | lsb_6 | 0.76661127 |
| 190 | lsb_2 | 0.750155875 | 190 | lsb_2 | 0.76394842 |
| 190 | lsb_1 | 0.650144744 | 190 | lsb_1 | 0.69899019 |
| 190 | msb_1 | 0.525655632 | 190 | msb_1 | 0.42515182 |

Figure A6 depicts the correlation results for the first identified CAN channel, related to the accelerator pedal, for the 2022 Chevrolet Traverse. The comparison between the recorded CAN data and the expected acceleration waveform demonstrates the system's effectiveness.

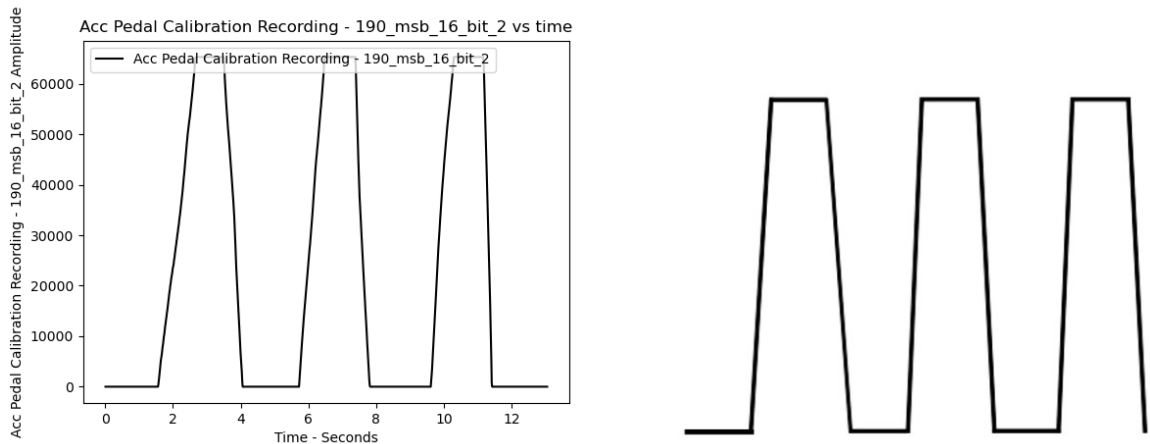

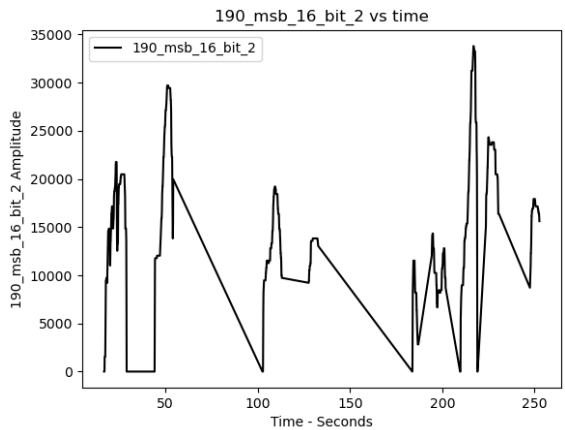
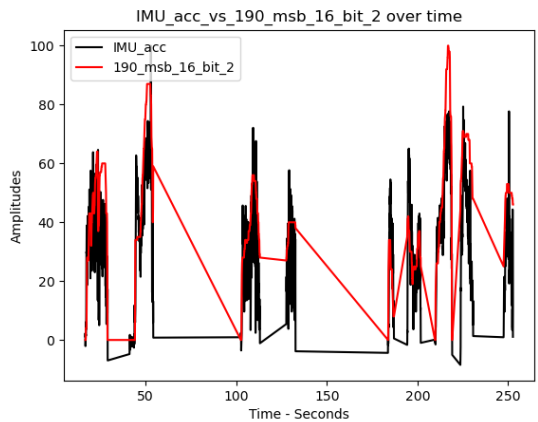

**Figure A6.** 2022 Chevrolet Traverse accelerator pedal validation one.

Figure A7 provides a similar visualization for a second identified CAN channel, related to the accelerator pedal, further demonstrating the accuracy of the system.

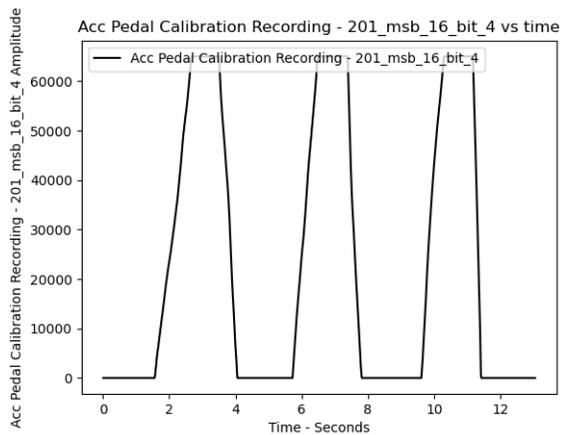
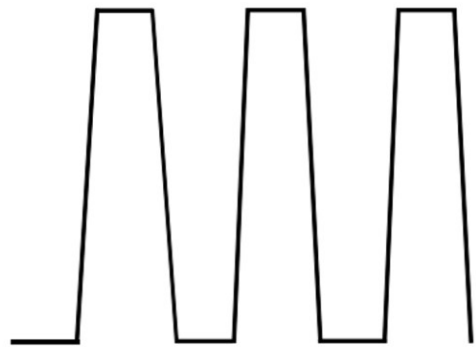
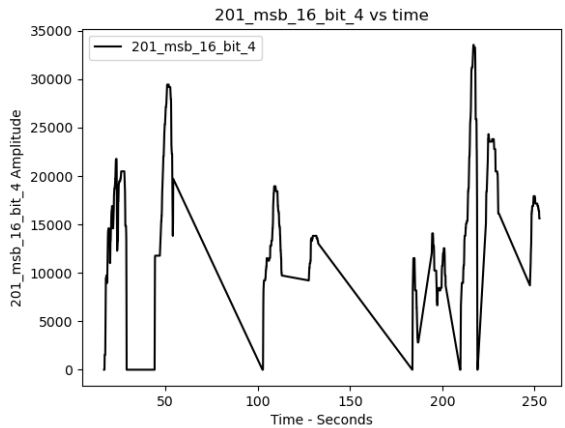
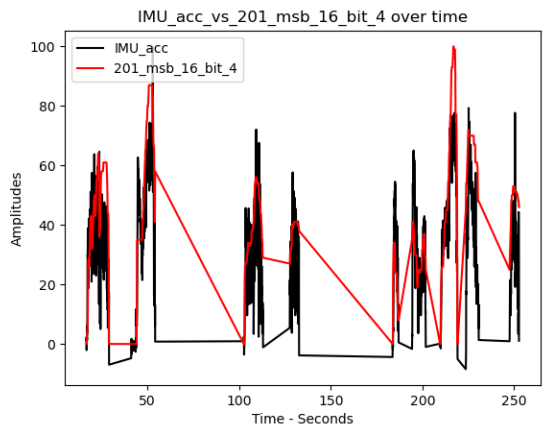

**Figure A7.** 2022 Chevrolet Traverse accelerator pedal validation one.

Table A7 presents the correlation data for the brake pedal, showing values for three and six identification events. The data indicates a high correlation for the identified CAN channels, demonstrating the system's ability to accurately reverse engineer brake pedal inputs.

**Table A7.** Deceleration results.

| Three Events | | | Six Events | | |
|---|---|---|---|---|---|
| ID | Channel | Correlation | ID | Channel | Correlation |
| 241 | msb_1 | 0.88829099 | 241 | msb_3 | 0.8791912 |
| 241 | msb_3 | 0.88829094 | 241 | msb_1 | 0.87919066 |
| 190 | msb_1 | 0.88718132 | 190 | msb_1 | 0.87815224 |
| 1417 | msb_0 | 0.49105805 | | | |

Table A8 shows the brake pedal data for nine identification events. It shows continued high correlation values, demonstrating the system's consistency.

**Table A8.** Deceleration results.

| 9 Events | | |
|---|---|---|
| ID | Channel | Correlation |
| 241 | msb_3 | 0.87184028 |
| 241 | msb_1 | 0.87183963 |
| 190 | msb_1 | 0.86594352 |

Figure A8 presents the time series data for the first identified brake pedal CAN channel. It illustrates the alignment of the channel's data with the expected brake pedal movements.

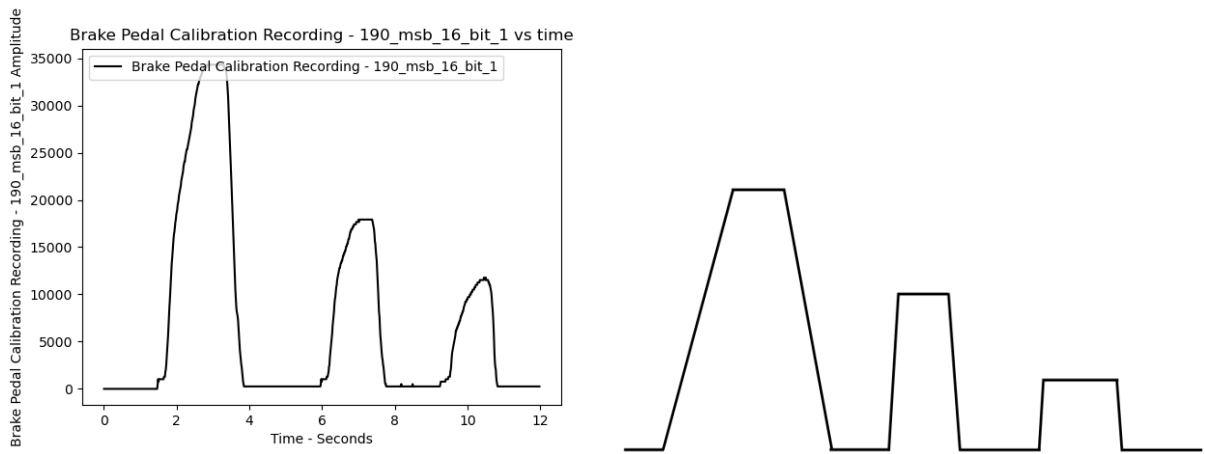

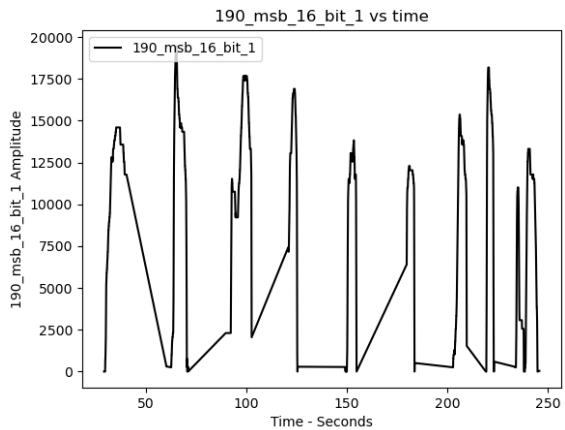
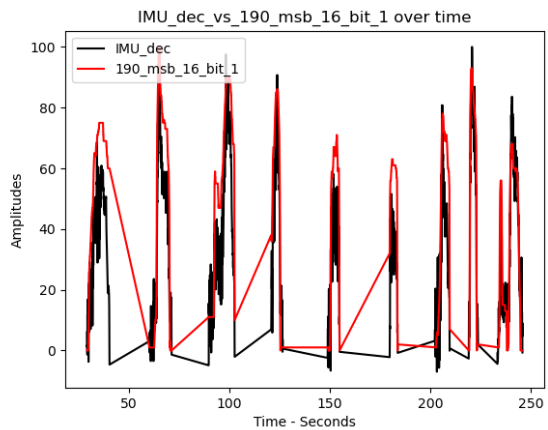

**Figure A8.** 2022 Chevrolet Traverse brake pedal validation one.

Figure A9 provides a similar time series visualization for the second identified brake pedal CAN channel, demonstrating the consistency of the system.

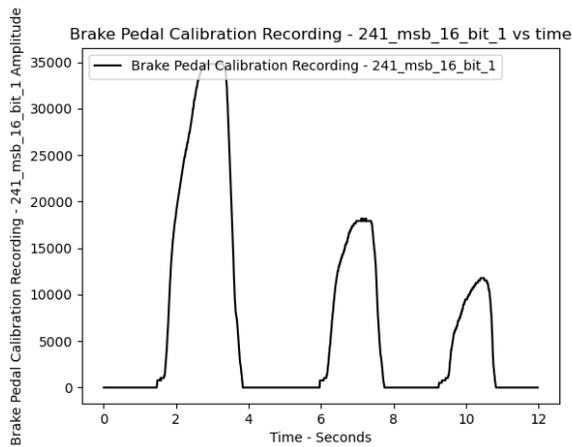
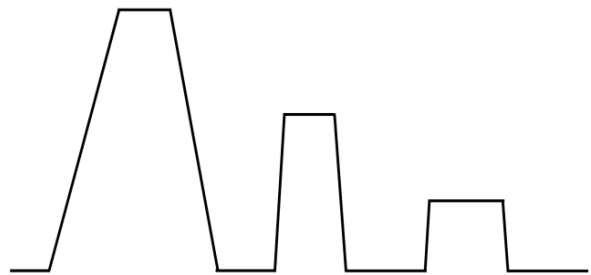
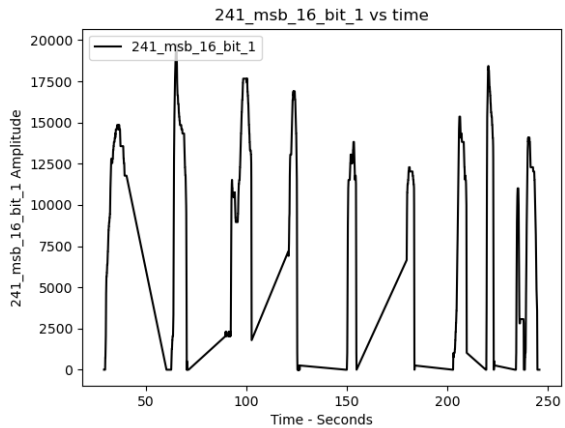
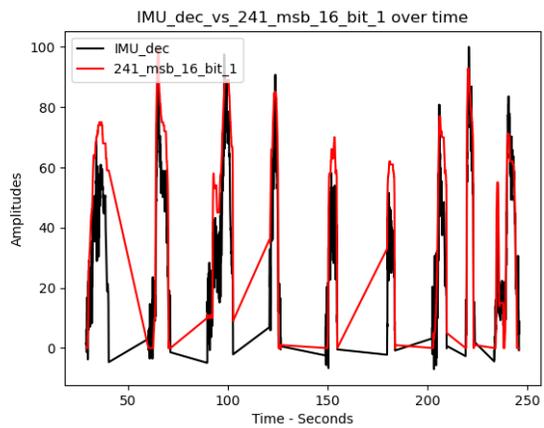

**Figure A9.** 2022 Chevrolet Traverse brake pedal validation two.

Table A9 shows the correlation values for the steering wheel CAN channel for three and six identification events. The data shows the system's ability to identify relevant channels with a limited set of events.

**Table A9.** Steering results.

| Three Events | | | Six Events | | |
|---|---|---|---|---|---|
| ID | Channel | Correlation | ID | Channel | Correlation |
| 564 | msb_2 | 0.498557714 | 564 | msb_2 | 0.5878403 |
| 426 | msb_4 | 0.430835564 | | | |

Table A10 presents the steering wheel CAN channel correlation value for nine events, demonstrating consistent identification accuracy.

**Table A10.** Steering results.

| Nine Events | | |
|---|---|---|
| ID | Channel | Correlation |
| 564 | msb_2 | 0.56611029 |

Figure A10 displays the time series data for the steering wheel CAN channel. It demonstrates the system's effectiveness in capturing and analyzing steering inputs.

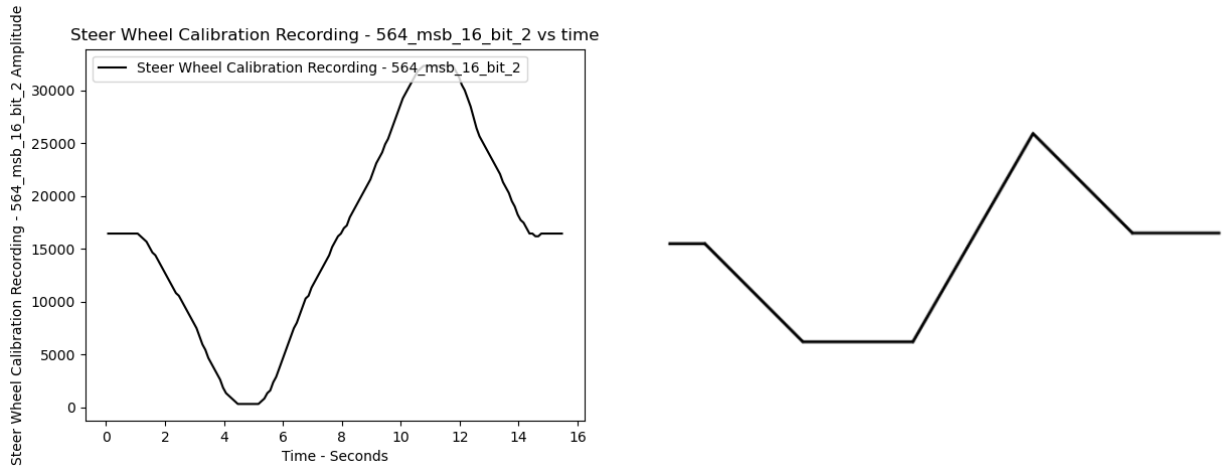

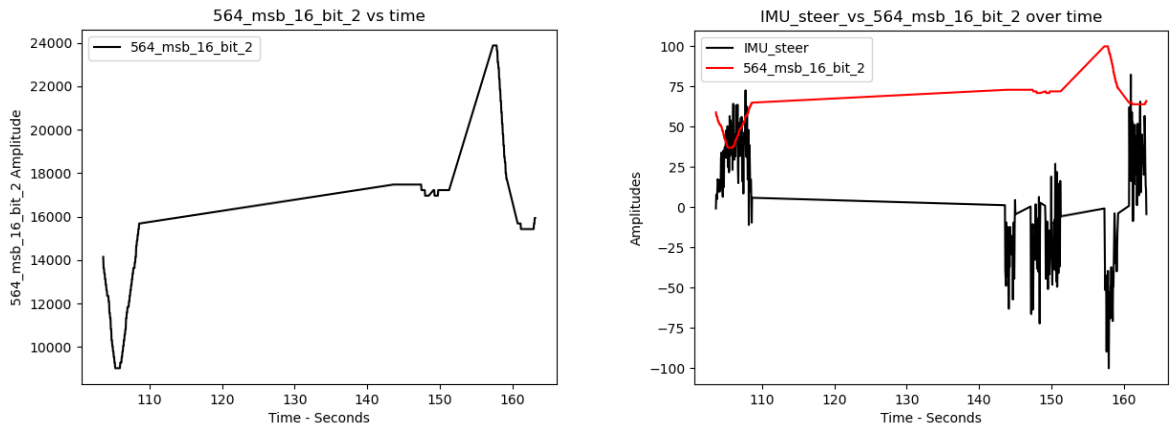

**Figure A10.** 2022 Chevrolet Traverse steering wheel validation.

*A.3. 2006 Volvo XC70 Results*

Results from the 2006 Volvo XC70 highlight the system's adaptability to older vehicle models. The identified CAN channels for this vehicle provided strong correlations, despite the age difference, as compared to the more modern vehicles tested.

Table A11 presents the correlation values for the CAN channels associated with the accelerator pedal for three and six identification events. Table A12 presents data for nine and twelve events. The high correlation values reflect the system's effectiveness in identifying relevant channels, even in older vehicles.

**Table A11.** Acceleration results.

| Three Events | | | Six Events | | |
|---|---|---|---|---|---|
| ID | Channel | Correlation | ID | Channel | Correlation |
| 6438942 | msb_6 | 0.733546351 | 6438942 | msb_6 | 0.71442166 |

**Table A12.** Acceleration results.

| Nine Events | | | Twelve Events | | |
|---|---|---|---|---|---|
| ID | Channel | Correlation | ID | Channel | Correlation |
| 6438942 | msb_6 | 0.733746815 | 6438942 | msb_6 | 0.78024858 |

Figure A11 depicts the correlation results for the identified CAN channel, related to the accelerator pedal, in the 2006 Volvo XC70.

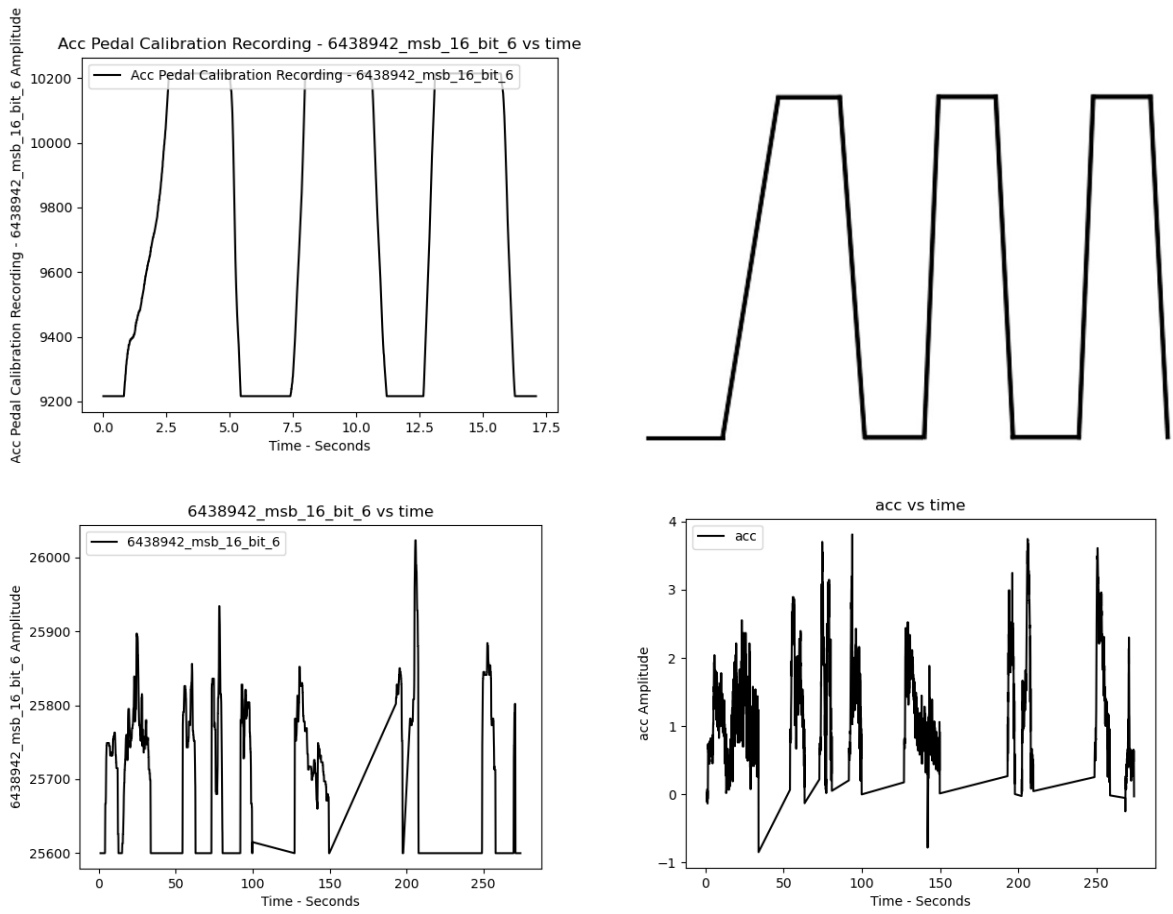
**Figure A11.** 2006 Volvo XC70 accelerator pedal validation.

Table A13 presents the results for the brake pedal. It shows strong correlation values for the identified CAN channel for three and six identification events.

**Table A13.** Deceleration results.

| Three Events | | | Six Events | | |
|---|---|---|---|---|---|
| ID | Channel | Correlation | ID | Channel | Correlation |
| 2244644 | msb_2 | 0.91161871 | 2244644 | msb_2 | 0.90388413 |

Figure A12 shows the time series for the brake pedal CAN channel. It illustrates how well the identified channel corresponds to the brake pedal's movement.

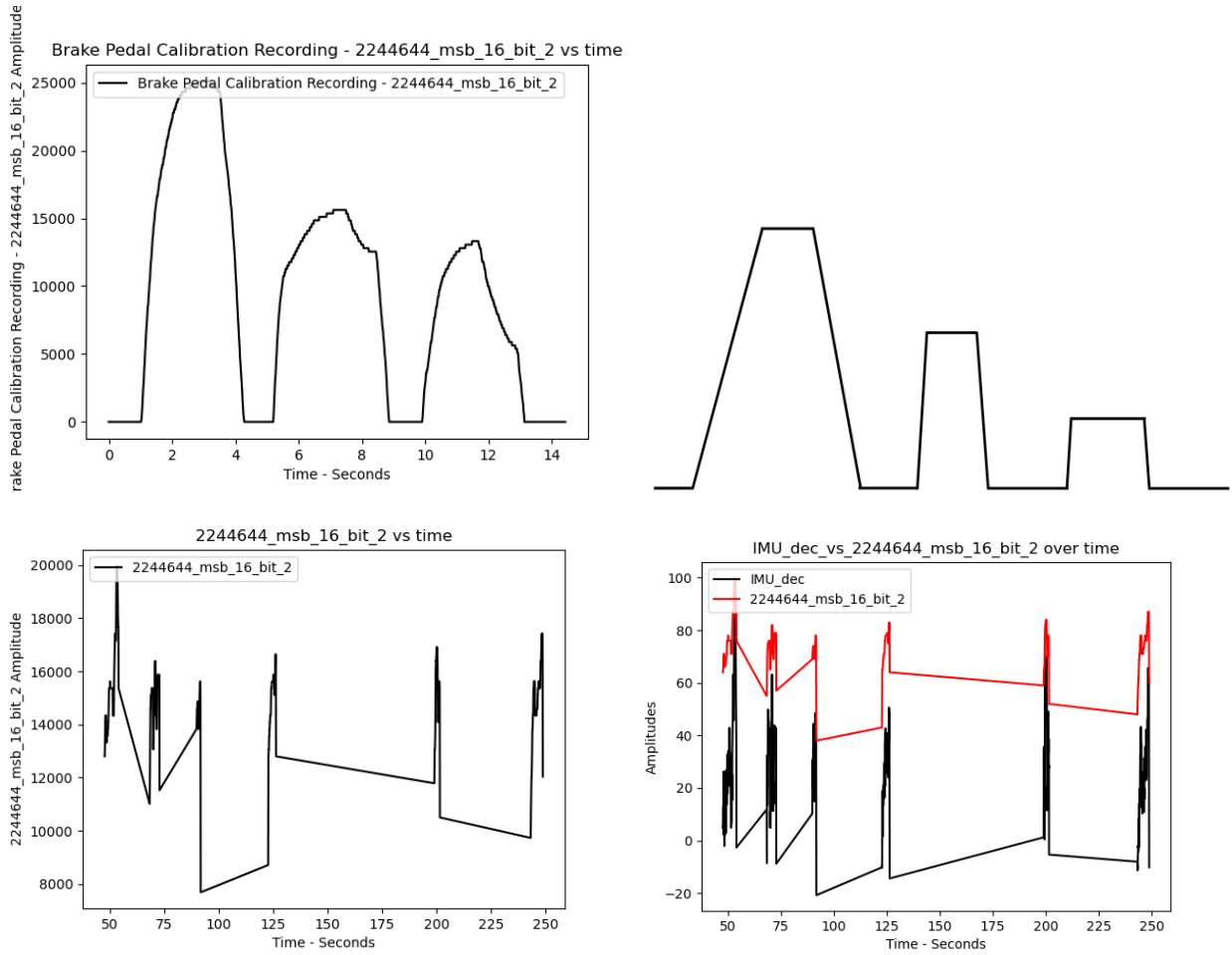

**Figure A12.** 2006 Volvo XC70 brake pedal validation.

Table A14 presents the correlation values for the steering wheel CAN channel for three and six identification events. It demonstrates the system's ability to accurately identify relevant channels in an older vehicle.

**Table A14.** Steering results.

| Three Events | | | Six Events | | |
|---|---|---|---|---|---|
| ID | Ch | Corr | ID | Ch | Corr |
| 283262976 | msb_2 | 0.769 | 283262976 | msb_2 | 0.7422 |

Table A15 presents the correlation value for the steering wheel CAN channel for nine events, showing the system's consistency in identifying CAN channels over time.

**Table A15.** Steering results.

| Nine Events | | |
|---|---|---|
| ID | Ch | Corr |
| 283262976 | msb_2 | 0.8054 |

Figure A13 presents the time series data for the steering wheel CAN channel, showing the system's capability to capture and analyze steering wheel inputs effectively in older vehicles like the 2006 Volvo XC70.

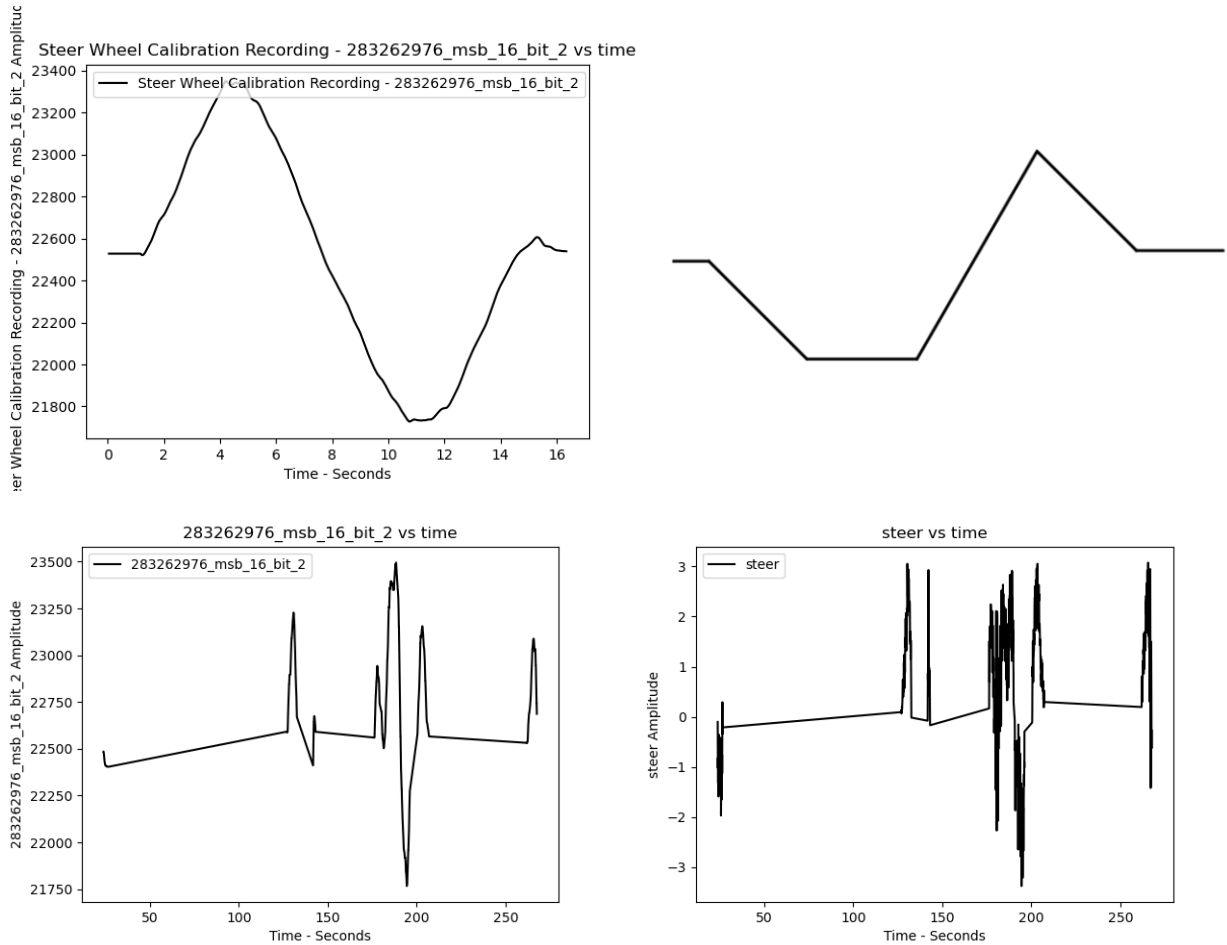

**Figure A13.** 2006 Volvo XC70 steering wheel validation.

*A.4. 2009 Chevrolet Impala Results*

The data from the 2009 Chevrolet Impala is particularly noteworthy, due to the challenges faced in identifying steering wheel CAN channels, as previously discussed. Despite these challenges, the system successfully identified CAN channels related to the accelerator and brake pedals.

Table A16 provides the correlation values for the identified CAN channels associated with the accelerator pedal over three and six events. The data indicates that the system was able to effectively identify relevant CAN channels for the accelerator pedal, for this older vehicle.

**Table A16.** Acceleration results.

| Three Events | | | Six Events | | |
|---|---|---|---|---|---|
| ID | Channel | Correlation | ID | Channel | Correlation |
| 201 | msb_4 | 0.72082 | 201 | msb_4 | 0.680740179 |

Figure A14 visualizes the results of the identified accelerator pedal CAN channel. It shows that the system successfully captured the accelerator pedal movements.

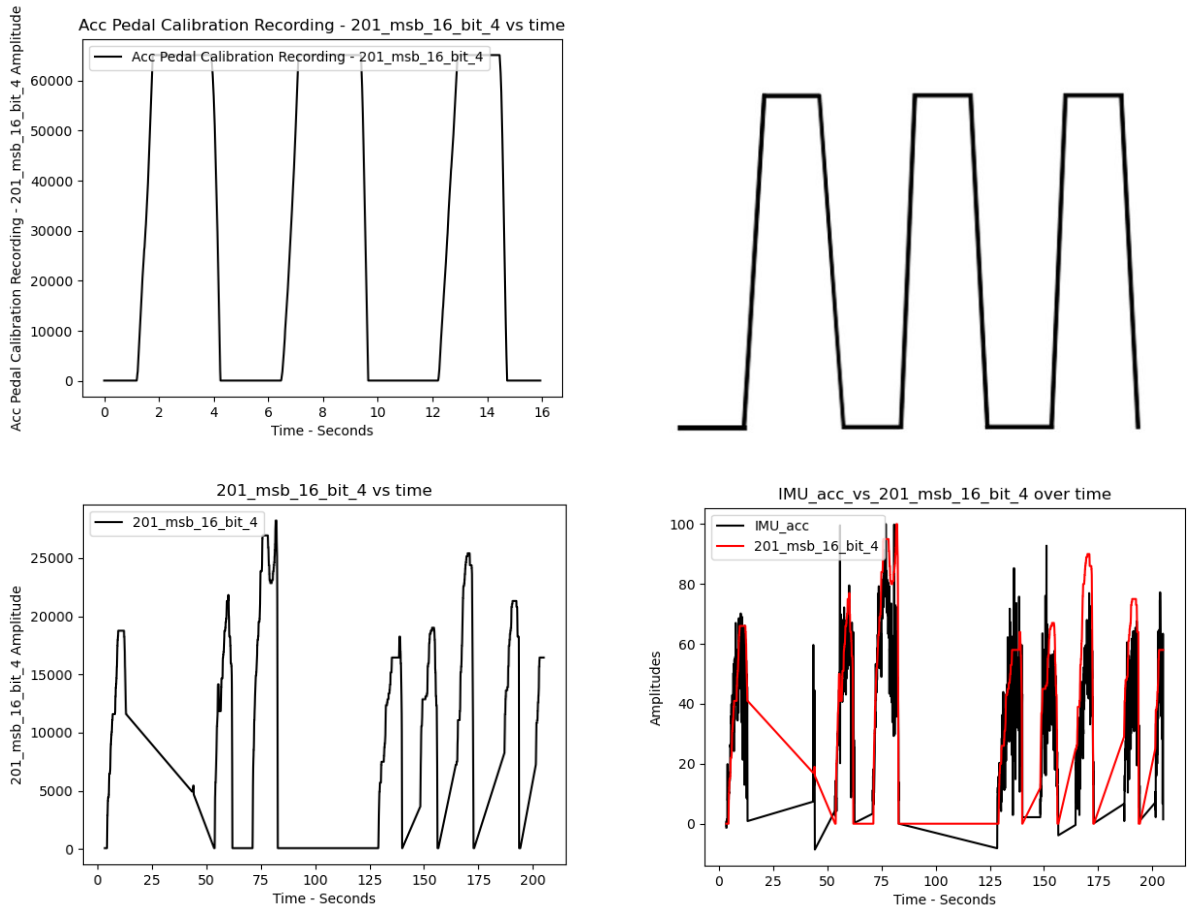

**Figure A14.** 2009 Chevrolet Impala accelerator pedal validation.

Table A17 shows the correlation values for the brake pedal CAN channels for three and six identification events. The results demonstrate that the system could accurately reverse engineer the brake pedal inputs for the 2009 Chevrolet Impala.

**Table A17.** Deceleration results.

| Three Events | | | Six Events | | |
|---|---|---|---|---|---|
| ID | Channel | Correlation | ID | Channel | Correlation |
| 241 | msb_1 | 0.793219665 | 241 | msb_1 | 0.78068928 |

Table A18 provides the correlation values for the brake pedal CAN channel for nine events, confirming the system's consistency.

**Table A18.** Deceleration results.

| Nine Events | | |
|---|---|---|
| ID | Channel | Correlation |

| 241 | msb_1 | 0.78157761 |
|---|---|---|

Figure A15 displays the time series data for the brake pedal CAN channel. It shows the correlation between the data and the brake pedal movements.

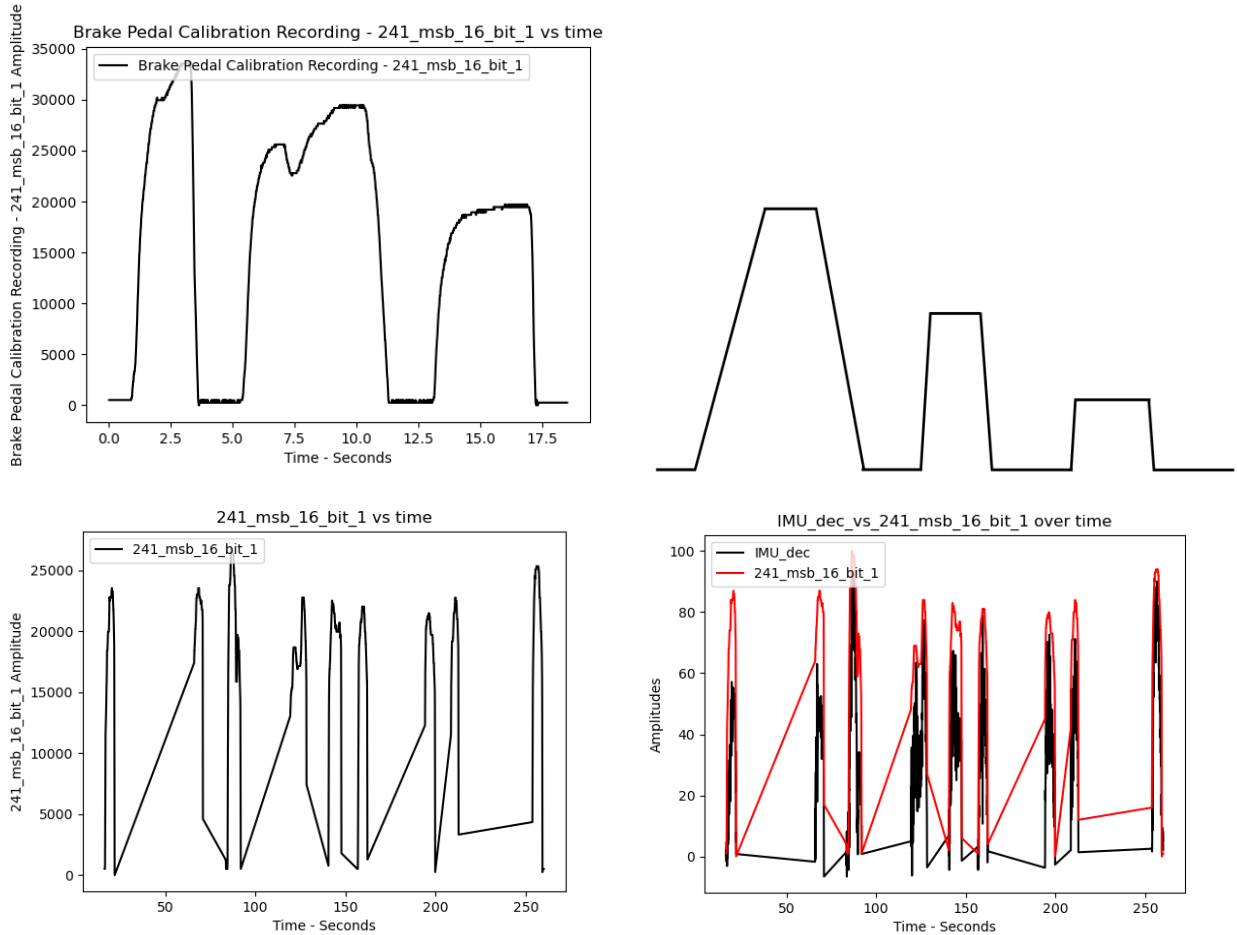

**Figure A15.** 2009 Chevrolet Impala brake pedal validation.

Table A19 presents the steering results for three and six identification events. Although the system identified some CAN channels, the steering wheel channel did not correlate as strongly as with other vehicles.

**Table A19.** Steering results.

| Three Events | | | Six Events | | |
|---|---|---|---|---|---|
| ID | Channel | Correlation | ID | Channel | Correlation |
| 707 | msb_1 | 0.5656046 | 455 | msb_0 | 0.5421081 |

Table A20 shows that, after nine events, the system was unable to converge on a specific CAN channel for the steering wheel. This was anticipated, based on prior research [5, 6]. This outcome shows the variation in CAN implementations across different vehicle models and years. It also demonstrates the system's ability to determine that a channel cannot be identified.

**Table A20.** Steering results.

| | Nine Events | |
|---|---|---|
| ID | Channel | Correlation |
| N/A | N/A | N/A |

As noted previously [6], the system was not expected to identify a strong correlation for the steering wheel channel in the 2009 Chevrolet Impala due to its CAN implementation.

### A5. 2016 Ford Fusion Results

The 2016 Ford Fusion was another modern vehicle tested. The results were consistent with those of other recent models, further demonstrating the system's effectiveness across different vehicle makes and models.

Tables A21 and A22 present the correlation values for the CAN channels associated with the accelerator pedal across three, six, and nine events. The results confirm the system's ability to identify relevant channels.

**Table A21.** Acceleration results

| Three Events (19 sec) | | | Six Events (29 sec) | | |
|---|---|---|---|---|---|
| ID | Channel | Correlation | ID | Channel | Correlation |
| 516 | msb_0 | 0.5791113 | 516 | msb_0 | 0.55247735 |

**Table A22.** Acceleration results.

| | Nine Events (35 sec) | |
|---|---|---|
| ID | Channel | Correlation |
| 516 | msb_0 | 0.55223308 |

Figure A16 illustrates the time series data for the accelerator pedal CAN channel. It shows how the system accurately captured the corresponding movements in the 2016 Ford Fusion.

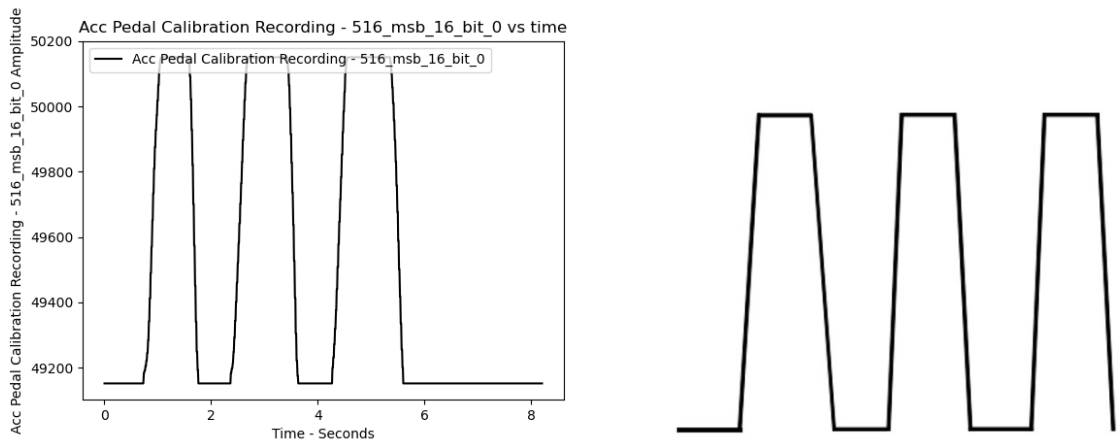

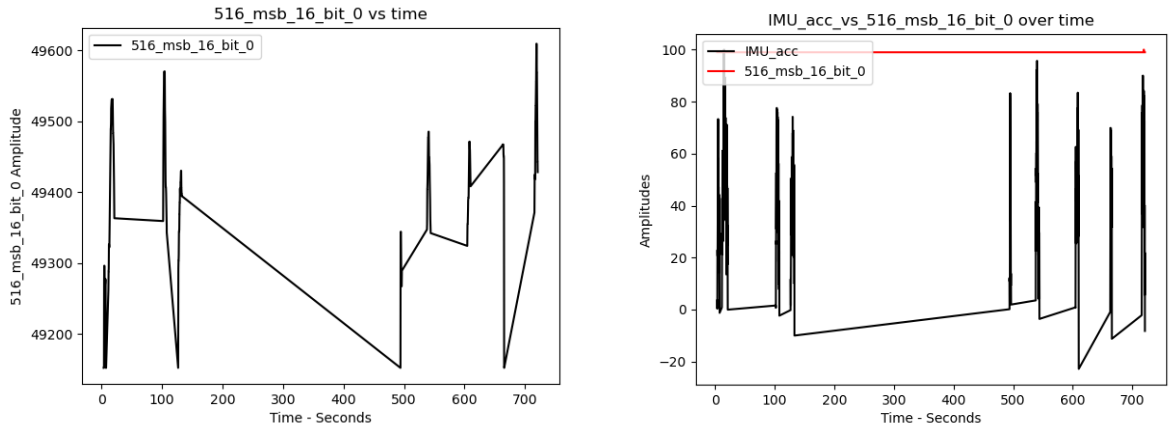

**Figure A16.** 2016 Ford Fusion accelerator pedal validation.

Table A23 presents the correlation values for the brake pedal CAN channels for three and six events. The data demonstrates the system's ability to reliably identify relevant channels related to brake inputs.

**Table A23.** Deceleration results

| Three Events (28 sec) | | | Six Events (40 sec) | | |
|---|---|---|---|---|---|
| ID | Channel | Correlation | ID | Channel | Correlation |
| 1200 | msb_5 | 0.9398591 | 1200 | msb_5 | 0.944226011 |
| 125 | msb_0 | 0.9377895 | 125 | msb_0 | 0.943019974 |

Table A24 shows the correlation values for the brake pedal CAN channels for nine identification events, showing the system's consistency.

**Table A24.** Deceleration results

| Nine Events (41 sec) | | |
|---|---|---|
| ID | Channel | Correlation |
| 1200 | msb_5 | 0.94459447 |
| 125 | msb_0 | 0.94385962 |

Figure A17 and Figure A18 present the time series data for the brake pedal CAN channels. This demonstrates the system's accuracy in real-time reverse engineering.

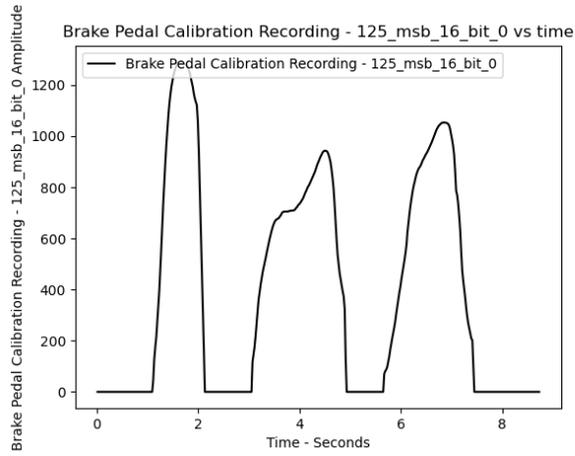 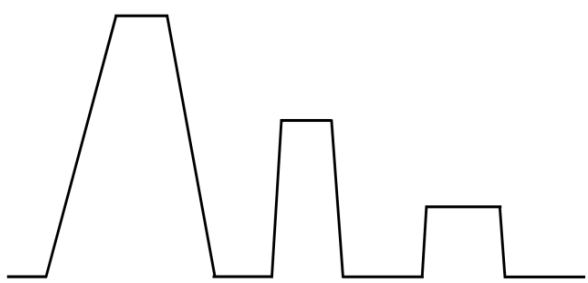
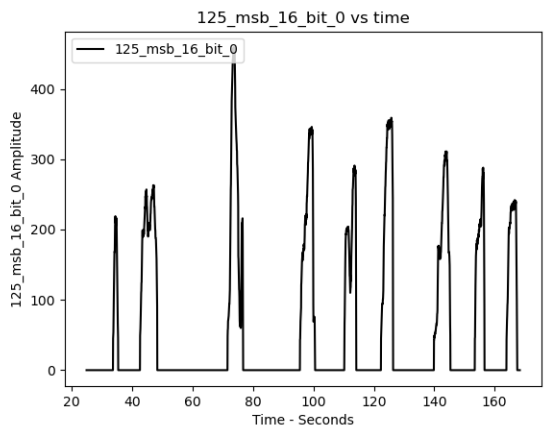 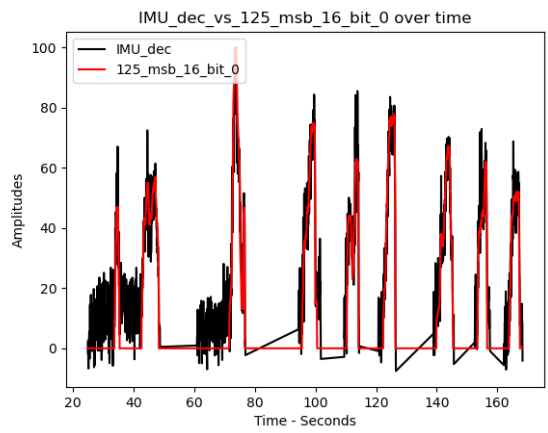

**Figure A17.** 2016 Ford Fusion brake pedal validation one.

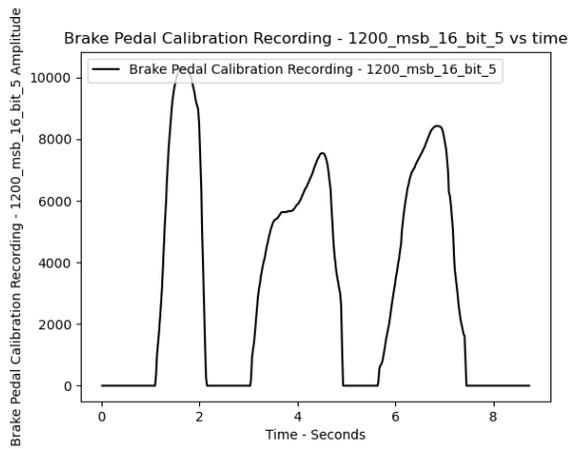 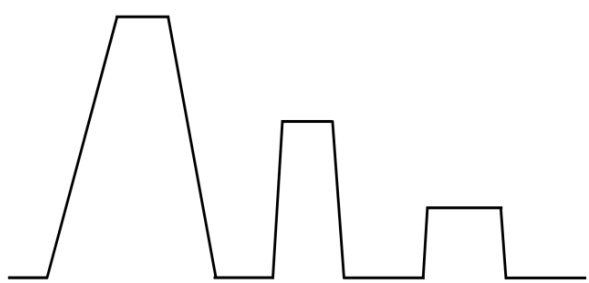

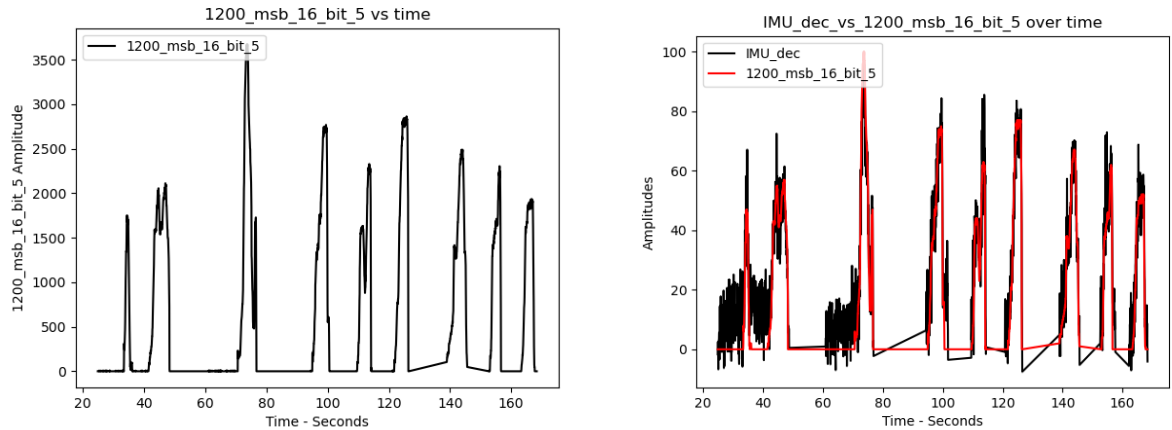

**Figure A18.** 2016 Ford Fusion brake pedal validation two.

Table A25 shows the correlation values for the steering wheel CAN channels for three and six identification events. The results indicate the system's consistent performance in identifying relevant channels for steering inputs.

**Table A25.** Steering results.

| Three Events (28 sec) | | | Six Events (42 sec) | | |
|---|---|---|---|---|---|
| ID | Channel | Correlation | ID | Channel | Correlation |
| 133 | msb_0 | 0.5466915 | 118 | msb_0 | 0.71991522 |
| 118 | msb_0 | 0.5460877 | 133 | msb_0 | 0.71886578 |

Table A26 presents the correlation values for the steering wheel CAN channels for nine identification events, further demonstrating the system's reliability.

**Table A26.** Steering results.

| Nine Events (59 sec) | | |
|---|---|---|
| ID | Channel | Correlation |
| 118 | msb_0 | 0.76039649 |
| 133 | msb_0 | 0.75951201 |

Figure A19 and Figure A20 display the time series for the steering wheel CAN channels. They demonstrate the system's effectiveness in capturing and analyzing steering inputs.

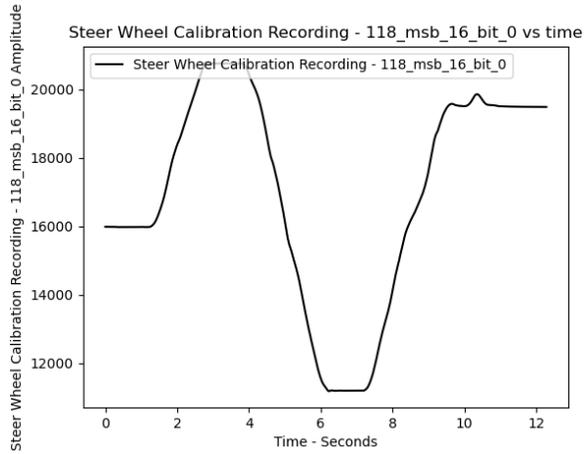
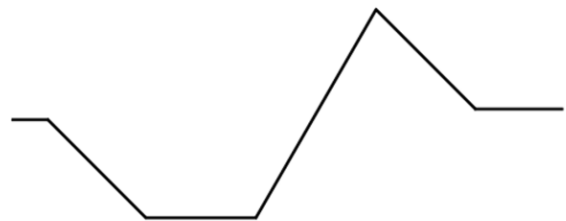
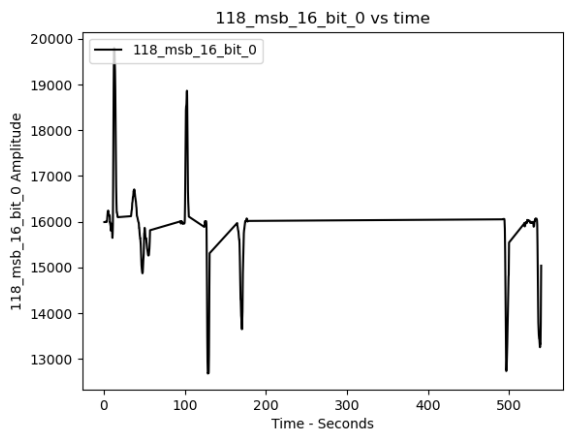
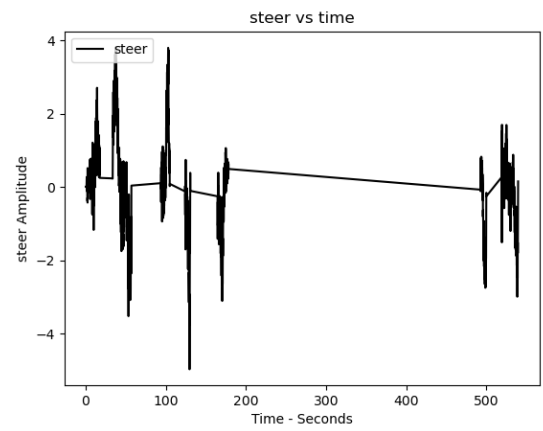

**Figure A19.** 2016 Ford Fusion steering wheel validation one.

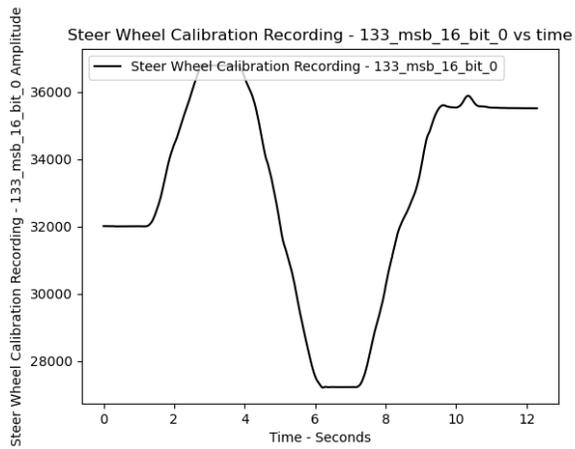
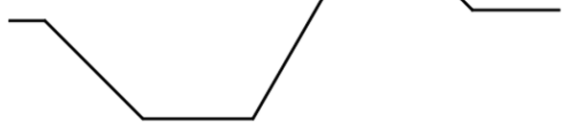

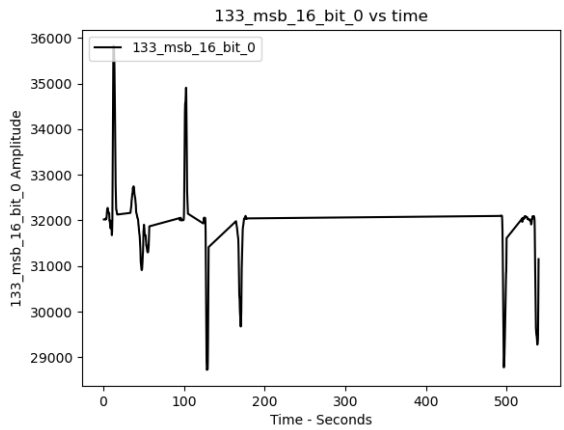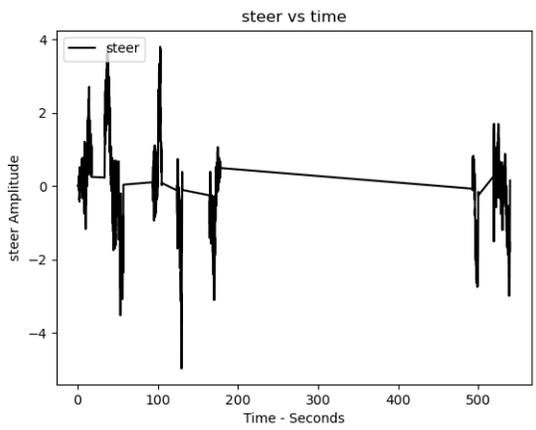

**Figure A20.** 2016 Ford Fusion steering wheel validation two.